\documentclass[fleqn,usenatbib]{mnras}
\usepackage{newtxtext,newtxmath}

\usepackage[T1]{fontenc}

\DeclareRobustCommand{\VAN}[3]{#2}
\let\VANthebibliography\thebibliography
\def\thebibliography{\DeclareRobustCommand{\VAN}[3]{##3}\VANthebibliography}

\usepackage{graphicx}   
\usepackage{amsmath}    
\usepackage{orcidlink}
\usepackage[flushleft]{threeparttable}
\usepackage{array}
\newcommand{\Rh}{{R_{\rm{Hm}}}}
\newcommand{\Rp}{{r_{\rm{p}}}}
\newcommand{\Mp}{{m_{\rm{p}}}}
\newcommand{\Np}{{N_{\rm{p}}}}
\newcommand{\Npi}{{N_{\rm{p,ini}}}}
\newcommand{\Pone}{P_{\rm{1}}}
\newcommand{\aone}{a_{\rm{1}}}
\newcommand{\imut}{i_{\rm{m}}}

\newcommand{\ecc}{{e}}
\newcommand{\inc}{{i}}

\newcommand{\Nsys}{{N_{\rm{sys}}}}
\newcommand{\porb}{{P_{\rm{orb}}}}
\newcommand{\kepler}{{{\em Kepler}}}

\newcommand{\cdpp}{{\rm{CDPP}_{\rm{4.5hr}}}}
\newcommand{\corr}{{\mathcal{C}}}

\title{Orbital Architectures of \kepler\ Multis From Dynamical Instabilities}

\author[T. Ghosh et al.]{
Tuhin Ghosh \orcidlink{0000-0002-3103-2000}$^{1}$\thanks{E-mail:\href{mailto:tghosh.astro@gmail.com}{tghosh.astro@gmail.com}}
and
Sourav Chatterjee \orcidlink{0000-0002-3680-2684}$^{1}$\thanks{E-mail: \href{souravchatterjee.tifr@gmail.com}{souravchatterjee.tifr@gmail.com}}\\
$^{1}$Department of Astronomy and Astrophysics, Tata Institute of Fundamental Research, Homi Bhabha Road, Colaba, Mumbai, 400005, India\\
}

\date{Accepted 2023 September 25. Received 2023 September 13; in original form 2023 April 24}
\pubyear{2023}

\begin{document}
\label{firstpage}
\pagerange{\pageref{firstpage}--\pageref{lastpage}}
\maketitle

\begin{abstract}
The high-multiplicity exoplanet systems are generally more tightly packed when compared to the solar system. Such compact multi-planet systems are often susceptible to dynamical instability. We investigate the impact of dynamical instability on the final orbital architectures of multi-planet systems using $N$-body simulations. Our models initially consist of six to ten planets placed randomly according to a power-law distribution of mutual Hill separations. We find that almost all of our model planetary systems go through at least one phase of dynamical instability, losing at least one planet. The orbital architecture, including the distributions of mutual Hill separations, planetary masses, orbital periods, and period ratios, of the transit-detectable model planetary systems closely resemble those of the multi-planet systems detected by \kepler. We find that without any formation-dependent input, a dynamically active past can naturally reproduce important observed trends including multiplicity-dependent eccentricity distribution, smaller eccentricities for larger planets, and intra-system uniformity. On the other hand, our transit-detectable planet populations lack the observed sub-population of eccentric single-transiting planets, pointing towards the `\kepler\ dichotomy'. These findings indicate that dynamical instabilities may have played a vital role in the final assembly of sub-Jovian planets.

\end{abstract}



%
\section{Introduction}\label{sec:intro}

Numerous successful space-based transit surveys such as NASA's \kepler\ \citep{2010Borucki, Borucki_2016}, {\em K2} \citep{Howell_2014, Cleve_2016}, and {\em TESS} \citep{2015RickerTESS}, numerous radial velocity surveys such as HIRES \citep{HIRES}, HARPS \citep{HARPS}, HARPS-N \citep{HARPS-N}, SOPHIE \citep{SOPHIE}, and ESPRESSO \citep{ESPRESSO} and follow-up missions like LAMOST-\kepler\ survey \citep{Lamost_Dong_2014, Lamost_Luo_2015} and California-\kepler\ Survey \citep{CKS}, have led to the discovery and characterization of more than $5000$ exoplanets \citep[NASA Exoplanet Archive, \footnote{http://exoplanetarchive.ipac.caltech.edu}][]{2013Akeson_NEA} in more than $3500$ planetary systems with diverse orbital architectures. Among these, NASA's \kepler\ mission stands out with over $2700$ confirmed planet discoveries, more than half of all planets discovered to date. The majority of these planets are smaller than Neptune and have relatively short orbital periods ($\porb/\rm{yr}\lesssim1$). Apart from the sheer number of exoplanet discoveries, the \kepler\ mission also found an abundance of multi-planet systems (multis in short). These multi-transiting systems not only inform us about the orbital architecture of planetary systems at present, but also provide key constraints on their formation, assembly, and dynamical evolution and they are also crucial for inferring the intrinsic exoplanet populations not directly observable due to selection biases \citep[e.g.,][]{2011_Lissauer, Fang_2012, 2014_Fabrycky, 2015PuWu, 2018_Weiss, 2019_He, 2020_He}.

Multi-planet systems are often tightly packed and susceptible to dynamical instabilities. Indeed, several studies have suggested that many observed multi-planet systems are very close to their stability limit \citep[e.g.,][]{Deck_2012, 2013Fang, 2013Lissauer, Volk_2015, Hwang2017, Volk_2020}. The timescale for the onset of dynamical instabilities in a planetary system is dependent on the spacing 
\begin{equation}
\label{eqn:hill_sep}
K = \frac{a_{j+1} - a_{j}}{\Rh}
\end{equation}
between the planets in multiples of the so-called mutual Hill radius
\begin{equation}
\label{eqn:hill_radius}
\Rh_{j,j+1} = \frac{a_{j} + a_{j+1}}{2} \left( \frac{\Mp_{j} + \Mp_{j+1}}{3M_{*}} \right)^{1/3}
\end{equation}
where $a_j$, $\Mp_j$ are the semi-major axis and mass of the j-th planet and $M_*$ is the mass of the host star. The stability of two planet systems can be analytically determined from this parameter \citep[e.g.,][]{GLADMAN1993, 2013Deck}. However, for systems with higher multiplicities, we rely on numerical simulations to estimate the stability timescale normalised by the innermost planet's $\porb$, $\tau$. Extensive numerical studies have revealed an empirical relation between $K$ and $\tau$, 
\begin{equation}
\label{eqn:stability_vs_K}
\log \tau \simeq bK + c
\end{equation}
where $b$ and $c$ are constants that depend on the system's multiplicity and planetary masses \citep[e.g.,][]{1996Chambers, 2010Funk}. Usually, to limit the parameter space, numerical investigations of $\tau$ often adopt equal mass planets separated at equal $K$. Even with these simplifications, $\tau$ shows substantial statistical variations, sometimes with ranges spanning orders of magnitude, and deviates from this relation near mean motion resonances \citep[e.g.,][]{Chatteerjee_2008,2010Funk}. How $\tau$ behaves for systems hosting planets of different masses and $K$ is not yet properly mapped. Qualitatively though, it is well understood that over time, the low-$K$ orbits interact and $K$ increases through collisions, ejections, and scattering. The systems, as a result, evolve towards longer $\tau$. Thus, the $K$-distribution of a population of multis essentially indicates a statistical measure of the dynamical state of the population.

Getting the spacing distribution for the observed multis is straightforward (\autoref{eqn:hill_sep}) if we know the semi-major axes (SMA) and planet masses ($\Mp$). However, $\Mp$ measurements are unavailable for a large fraction of \kepler\ planets. If we circumvent this challenge by using a mass-radius relationship \citep[e.g.,][]{Chen_2016} to estimate $\Mp$ from the measured radii ($\Rp$), the $K$-distribution for \kepler\ multis peaks around $K\sim12-15$, expected for a stability timescale of $\gtrsim$billion years \citep{2015PuWu}. There is a sharp drop in the number for $K\lesssim12$ and a power-law-like drop-off for higher $K$. Based on this, it has been suggested that the present-day $K$ distribution may be a result of past dynamical sculpting \citep[e.g.,][]{2015PuWu, Volk_2015}. Past dynamical sculpting not only affects the distribution of $K$ but also leaves tell-tale signatures in the orbital properties including eccentricities ($\ecc$) and mutual inclinations ($\imut$) \citep[e.g.,][]{Rasio_Ford_1996, Chatteerjee_2008, 2008Juric_Tremaine, Nagasawa_2011}.
Several studies investigated the effects of planetary dynamics in planetary systems with a variety of initial conditions, usually motivated by a preferred formation scenario. For example, \citet{2017Izidoro, 2021_Izidoro, Goldberg_2022} studied the effects of dynamical instability on systems of super-earths, initially in compact resonant chains. \citet{2019Dong-Hong_Wu} investigated the influence of dynamical instability on an initially flat period ratio distribution.

In contrast, in this study we are intentionally agnostic towards any particular formation scenario and make simple assumptions about the initial conditions of planetary systems as they start to evolve freely without any gas disk. We want to investigate, if we do not provide any input from formation scenarios or inject any intra-system trends, how close we can get to the observed properties and correlations for the \kepler\ multis simply as a result of dynamical instabilities. Since the majority of the \kepler\ planets are smaller and the larger giant planets are expected to form differently, we focus on planetary systems containing sub-Jovian ($\Rp\lesssim6R_{\oplus}$) planets only in this paper. In \autoref{sec:setup}, we describe the numerical setup and the adopted initial conditions. In \autoref{sec:results}, we present the outcomes from our suite of simulations and compare our models with \kepler's observations after correcting for transit geometry and \kepler's detection efficiency. Finally, we outline the key results and conclude in \autoref{sec:summary}.

\section{Numerical Setup} \label{sec:setup}

We construct $N$-body models of planetary systems with planets randomly assigned according to simple assumptions on the initial conditions without any input on intra-system features or correlations. Our fiducial ensemble initially consists of 8 planets in each system following \kepler-90, the system with the highest number of confirmed exoplanets \citep{2018_Kepler-90}. However, we create other ensembles by varying the initial number of planets ($\Npi$) to investigate the effects of $\Npi$ on our results. Furthermore, we explore the effects of initial $\ecc$ and inclination ($\inc$) distributions on our results. Below we describe the details of how we construct our initial planetary systems in different ensembles.

\subsection{Star and Planet Properties} \label{subsec:setup/star-planet-properties}
The central star in each model planetary system is chosen randomly from the planet-hosting stars discovered by \kepler. We draw $\Rp$ randomly from a power law distribution $f(\Rp)\propto\Rp^{\alpha_r}$ with $\alpha_r=-1.08$ based on the estimated intrinsic $\Rp$ distribution \citep{2019_He} in the range $0.5\leq\Rp/R_\oplus\leq3$. The occurrence rate of sub-jovian planets drops significantly for $\Rp/R_\oplus\gtrsim3$ \citep[e.g.,][]{Petigura_2013}. On the other hand, the detection probability of planets smaller than $\Rp/R_\oplus\lesssim 0.5$) is low. We estimate $\Mp$ from $\Rp$ using the probabilistic relationship given in \citet{Chen_2016}. We randomly draw $\Mp$ within $1\sigma$ of the prescribed normal distribution for a given $\Rp$. These choices ensure that all planets in our ensembles are initially sub-Neptunes and have $\Mp/M_{\earth} \lesssim 16$.

\subsection{Interplanetary separations \& initial planetary orbits} \label{subsec:setup/orbits}

Depending on the focus, past studies adopted various schemes for spacing the initial planets, such as a Gaussian distribution centering a fixed value \citep{2015PuWu} and a flat distribution in period ratios \citep{2019Dong-Hong_Wu}. In contrast, we space our planets using a $K$ distribution motivated by that for the \kepler\ multis and the following considerations.

The observed $K$ distribution shows a sharp decline for $K\lesssim12$ and a power-law like decline for $K\gtrsim15$. \citet{2015PuWu} suggested that the latter is genuine and not a result of observational biases. Planetary orbits with low separations ($K\lesssim12$) are expected to become unstable within the typical age of the observed host stars and hence, the sharp decline in low $K$ may be due to past dynamical instabilities. On the other hand, orbits separated by high $K\gtrsim15$ likely remained stable. Thus, the high end of the present-day $K$ distribution is expected to be more representative of the initial. Hence, we adopt a power-law distribution for the initial $K$ of the form $f(K)\propto K^{\alpha_K}$ in the range $K=4$ to $40$. The lower limit is slightly higher than the analytic limit for the stability of two-planet systems \citep{GLADMAN1993}. The upper limit comes from practical considerations. While the tail can extend to very high $K$, we truncate at $K=40$ to limit the physical size of our initial model systems. Planets in such high-SMA orbits are usually significantly beyond the detection capability of the \kepler\ mission, the focus of our study. Moreover, the fraction of observed planet pairs with $K>40$ is low ($<14\%$). 

Determining $\alpha_K$, is not straightforward. One naive choice could be to fit the observed $K$ distribution for large-$K$ to estimate $\alpha_K$. However, there is a significant caveat. A planetary system with low initial $K$ pairs may dynamically evolve to create high-$K$ pairs and contribute to the present-day $K\gtrsim15$. Hence, simply finding the best-fit power-law from the high end of the present-day $K$ distribution may not be justified. We want to find out the initial $K$ distribution that would result in an observed present-day distribution after dynamical evolution considering detection biases. Hence, we run several test simulations with different $\alpha_K$. Based on these tests we adopt $\alpha_K=-1.1$ throughout this study.

\begin{table}
    \centering
    \caption{Initial properties of the simulated ensembles and the corresponding $t_{\rm{stop}}$ (\autoref{subsec:setup/stopping}). Ensemble names contain $\Npi$, $\bar{e}$, and $\bar{\inc}$ for easy understanding of the corresponding initial properties. }
    \label{tab:initial_props}
    \begin{tabular}{cccccc}
        \hline
            \hline
            Ensemble & $\Nsys$ &  & Initial  & & $t_{\rm{stop}}/\Pone$\\
            \cline{3-5}
        Name &  & $\Npi$ & $\bar{\ecc}$ & $\bar{\inc}$ & \\
        \hline
            n8-e040-i024* & 355 & 8  & $0.040$ & $0.024$  &  $10^{8}$ \\
            n8-e025-i024  & 357 & 8  & $0.025$ & $0.024$  &  $10^{9}$ \\
            n8-e030-i024  & 356 & 8  & $0.030$ & $0.024$  &  $10^{9}$ \\
            n8-e035-i024  & 355 & 8  & $0.035$ & $0.024$  &  $10^{9}$ \\
            n8-e050-i024  & 356 & 8  & $0.050$ & $0.024$  &  $10^{8}$ \\
            n8-e060-i024  & 357 & 8  & $0.060$ & $0.024$  &  $5\times10^{6}$ \\
            n8-e080-i024  & 352 & 8  & $0.080$ & $0.024$  &  $6\times10^{5}$ \\
            n8-e040-i010  & 353 & 8  & $0.040$ & $0.010$  &  $10^{9}$ \\
            n8-e040-i050  & 353 & 8  & $0.040$ & $0.050$  &  $10^{7}$ \\
            n8-e040-i075  & 354 & 8  & $0.040$ & $0.075$  &  $5\times10^{6}$ \\
            n6-e040-i024  & 357 & 6  & $0.040$ & $0.024$  &  $10^{9}$ \\  
           n10-e040-i024  & 350 & 10 & $0.040$ & $0.024$  &  $5\times10^{7}$ \\
        \hline
    \end{tabular}
\end{table}

For each planetary system, we assign the SMA of the innermost planet ($\aone$) by randomly choosing (with replacement) from the observed $\aone$ in \kepler\ multis.
We place the remaining $\Npi-1$ planets in orbit one by one from inside out by randomly drawing $K_{\rm{rand}}$ from the adopted $K$ distribution. The SMA of the $(j+1)$th planet is then given by   
\begin{equation}
\label{eqn:a-calc}
a_{j+1} = a_{j} \frac{1 + X_{j}}{1 - X_{j}}
\end{equation}
where, 
\begin{equation}
\label{eqn:X-calc}
X_{j} = \frac{1}{2} K_{\rm{rand,j}} \left (\frac{\Mp_{j} + \Mp_{j+1}}{3M_{*}} \right)^{1/3}.
\end{equation}
We repeat this process until the SMA of the outermost planet is calculated. Note that our initial assembly does not specifically put any planet pair in or out of resonance. In effect, without any dissipation in the system, they are all initially non-resonant. For each orbit, orbital $\ecc$ and $\inc$ are drawn randomly from Rayleigh distributions with adopted mean values $\bar{\ecc}$, $\bar{\inc}$. We create several ensembles by varying $\Npi$, $\bar{e}$, and $\bar{\inc}$ (summarised in \autoref{tab:initial_props}). Other orbital elements, such as the longitude of ascending node, the argument of pericenter, and true anomaly, are chosen uniformly in their respective full ranges.

\subsection{Evolving the Planetary Orbits} \label{subsec:setup/evol}
We use the \texttt{mercurius} hybrid integration scheme \citep{2019REBMERCURIUS} implemented in the \texttt{rebound} simulation package \citep{ReinREB2012} to evolve the planetary systems with initial time step $\delta t_i=1/25$ of the innermost planet's $\porb$. We resolve close encounters by switching from the \texttt{WHFast} to the \texttt{IAS15} integrator when the planet-planet or planet-star separations are $3$ times the sum of their Hill radii. The typical energy error in our simulations is $\delta E/E\sim10^{-6}$, after correcting for energy losses during collisions and ejections. For each ensemble, we simulate $360$ planetary systems. We find that energy was not conserved well ($\delta E/E>10^{-3}$) in $\lesssim 2\%$ of these simulations, so we discard them from our analysis. \autoref{tab:initial_props} lists the number of systems that satisfy the minimum accuracy for energy conservation and are considered for further analysis. We treat collisions adopting the ``sticky-sphere" approximation- whenever two planets physically touch each other, we merge them conserving mass, linear momentum, and physical volume. We remove planets if the SMA is $>100$ times the initial outermost planet's initial $a$ to reduce computational cost and consider it an ejection (happened in case of $\lesssim0.7\%$ simulated planets).

\subsection{Transit Probability and \kepler's Detection Bias} \label{subsec:setup/detections}
The $N$-body simulations provide us with an intrinsic set of planetary systems sculpted through dynamical instability. To compare these planetary systems with those observed, we account for the geometric transit probability and \kepler's detection biases.

We use the \texttt{CORBITS} algorithm \citep{CORBITS_Brakensiek_2016} to identify the planets that would transit when viewed from a random line of sight (LOS). For each planetary system, we collect $1000$ LOSs for which at least one transiting planet is found.\footnote{Since we already choose $\aone$ from the observed sample, the bias in transit probability towards close-in planets is already imprinted into our sample. Hence, we use equal numbers of successful LOSs for all planetary systems.} Even if a planet transits, detection is not guaranteed due to sensitivity biases inherent in \kepler's pipeline. We adopt a global detection efficiency model for \kepler\ \citep{Burke_2015, 2015_Christiansen, 2016_Christiansen, 2017_Christiansen} to estimate the detection probability ($p_{\rm{det}}$) in the \kepler\ pipeline based on the signal to noise ratio (SNR) as a proxy to the so-called multi-event statistic (MES). We estimate SNR based on the combined differential photometric precision \citep[$\cdpp$,][]{2012_Christiansen} for 4.5hr duration of the host star as 
\begin{equation}
\label{eqn:snr}
    \rm{SNR} = \frac{(\Rp/R_{*})^2}{\cdpp \sqrt{4.5hr/T_{\rm{dur}}}}\sqrt{\frac{t_{\rm{obs}} * f}{\porb}},
\end{equation}
where $R_{*}$ is the stellar radius, $t_{\rm{obs}}$ is the time span of \kepler's observation, $f$ is the fraction of $t_{\rm{obs}}$ that contain valid data, and $T_{\rm{dur}}$ is the transit duration of the planet. We estimate $T_{\rm{dur}}$ using equation $1$ of \citet{Burke_2015}.

We estimate $p_{\rm{det}}$ in the \kepler\ pipeline using the gamma cumulative probability distribution given in \citet{2017_Christiansen}. We multiply $p_{\rm{det}}$ with an analytic window function \citep[][their Equation\ 6]{Burke_2015} which estimates the probability of detecting at least three transits, the minimum number adopted by \kepler\ to be considered detected, accounting for the limits of data coverage. We consider a transiting planet to be detected by \kepler\ if a random number from a uniform distribution $\mathcal{U}[0,1]$ is less than the planet's $p_{\rm{det}}$. Finally, from the intrinsic population, we find a population that would be detected by \kepler\ and call this population `model-detected' henceforth. Since we explicitly use \kepler's detection pipeline, we restrict ourselves to confirmed multis detected by \kepler\ for a fair comparison.

\subsection{Integration Stopping Criteria} \label{subsec:setup/stopping}

It is usually computationally impractical to continue $N$-body integration till the physical age of a typical system, neither can $N>2$ systems be proven stable. Traditionally, the adopted stopping time in $N$-body simulations are somewhat ad hoc and guided by practical considerations. Any dynamically active system becomes more and more stable through repeated dynamical encounters by decreasing $\Np$ or by increasing $K$ or both. Hence, the integration is usually stopped after some predefined number of orbits when scattering encounters become sufficiently rare. The challenge is that since the dynamically active system, over time moves towards increased stability and the stability timescale increases exponentially with increasing $K$, it is not clear what can be considered as sufficiently rare. 

We adopt a more sophisticated and physically meaningful approach to determine the integration stopping time. This is above and beyond the usually adopted criteria of `sufficiently' rare subsequent interactions (see \autoref{appendix:stopping} for more details). Since our testing hypothesis is that exoplanet systems formed more dynamically packed and we observe them after long dynamical sculpting, we want to bring our model population to a similar dynamical state as \kepler's multis. Since for any system, dynamical stability is determined by $K$, the observed $K$ distribution should represent the present-day dynamical state of the observed population.\footnote{Paying attention to the dynamical state of collisional $N$-body systems and {\em not} their physical ages when dynamical properties are of interest, is quite common in star cluster studies. The relavant timescale there is the two-body relaxation time \citep{HeggieHut2003}. In case of multi-planet systems, the equivalent timescale is the timescale for instability.} However, the observed $K$ distribution is not intrinsic and subject to geometric transit probability and pipeline detection efficiency. Hence, we must compare the synthetic $K$ distribution after taking into account these observational biases. We achieve this in the following way. 

We simulate all ensembles to at least $10^{8}\Pone$ and examine the model-detected $K$ distribution. If the model-detected $K$ distribution peaks at a higher value than the observed, the ensemble is too dynamically evolved compared to the observed population. If so, we look back in time to find a snapshot where these two distributions are statistically identical. Similarly, if the model-detected $K$ distribution peaks at a lower value, the ensemble has not yet reached the dynamical state of the observed population. We integrate these ensembles further until agreement is achieved or we reach our hard stop at $10^9\Pone$ (to limit computational cost). The time taken to achieve the desired population-level dynamical state $t_{\rm{stop}}$ depends on the distributions of initial orbital properties of the ensemble (\autoref{tab:initial_props}). Only ensemble \texttt{n6-e040-i024}, with low initial $\Npi$ does not quite reach the observed dynamical state at $10^9\Pone$.

\section{Results} \label{sec:results}

We first discuss the results from our fiducial ensemble (\texttt{n8-e040-i024}) and investigate its similarity with \kepler's multis. Afterward, we discuss the effects of different initial conditions on the final outcomes of our simulations.

\subsection{Fiducial Ensemble} \label{subsec:res/fiducial_model}
We find that about $97 \%$ of our synthetic planetary systems are affected by dynamical instabilities that result in the loss of at least one planet via collision with another planet or the host star or ejection from the system. Since our ensemble is dominated by close-in planetary systems with Safronov number $\Theta\ll1$, strong scattering ultimately results in collisions rather than ejections in most cases.\footnote{$\Theta$ is the squared ratio of the escape velocity from a planet’s surface to that from the planetary system at the location of the planet, \cite{Safronov}).}

\subsubsection{Orbital and Planetary Properties} \label{subsec:res/orbital_properties}
\begin{figure}
\includegraphics[width=\columnwidth]{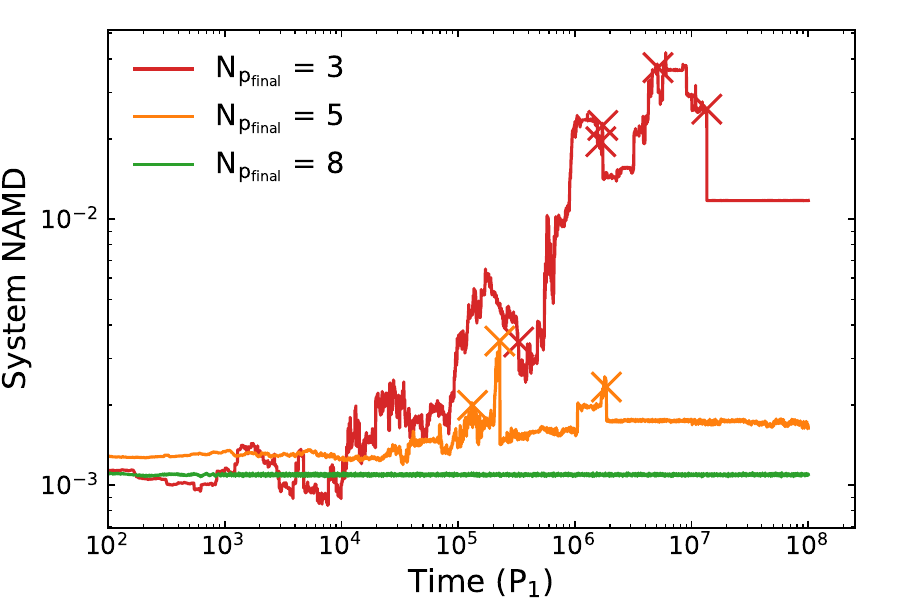}
\caption{Evolution of the NAMD for three different example systems. Although all systems initially had roughly the same NAMD, they follow different dynamical evolution with different final $\Np$ and NAMD. The crosses denote planet-planet collisions in each system.}
\label{fig:namd_evol}
\end{figure}

The final systems emerging from this chaotic dynamical evolution have little memory of their initial conditions. Hence, the orbital architectures of these systems are very different from their initial configuration. By design, our models initially have more planets at closer separations, and systems with such small interplanetary spacing inevitably face dynamical instabilities. Dynamical instability leads to close encounters between planets, increasing their dynamical excitation. The degree of dynamical excitation of a planetary system can be quantified by the normalised angular momentum deficit \citep{2001_Chambers},
\begin{equation}
\label{eqn:namd}
{\rm NAMD} = \frac{\sum_{j} {\Mp_{j}}\sqrt{a_{j}}\left(1-\sqrt{1-\ecc_{j}^{2}} \cos \inc_{j} \right)}{\sum_{j} {\Mp_{j}}\sqrt{a_{j}}}.
\end{equation}
We show the evolution of NAMD for three different representative systems in \autoref{fig:namd_evol}. In general, dynamical instabilities raise the NAMD of the systems while physical collisions decrease NAMD \citep[also see,][]{2001_Chambers, 2017_Laskar}. Competition between these two processes shapes the final interplanetary spacing and orbital properties in a system \citep{2016_Dawson}. On the other hand, NAMD remains conserved for systems that do not experience significant instability.

\begin{figure}
\includegraphics[width=\columnwidth]{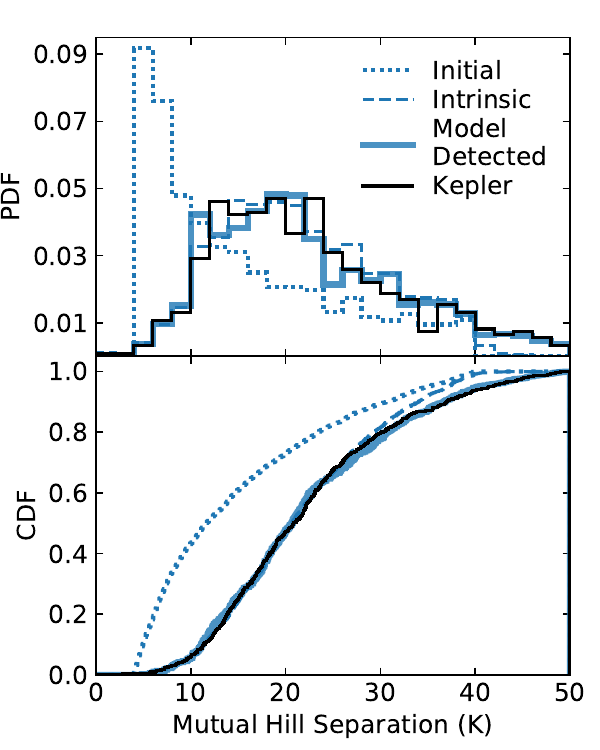}
\caption{Probability distribution function (PDF, top) and cumulative density function (CDF, bottom) for the mutual Hill separations ($K$) in the fiducial ensemble (\texttt{n8-e040-i024}). Black denotes \kepler's multis. The dotted, dashed, and solid blue lines denote the initial, final intrinsic, and final model-detected populations. The model-detected and observed $K$ distributions are in excellent agreement. 
}
\label{fig:K_dist_fidu}
\end{figure}
\begin{figure*}
\includegraphics[width=\textwidth]{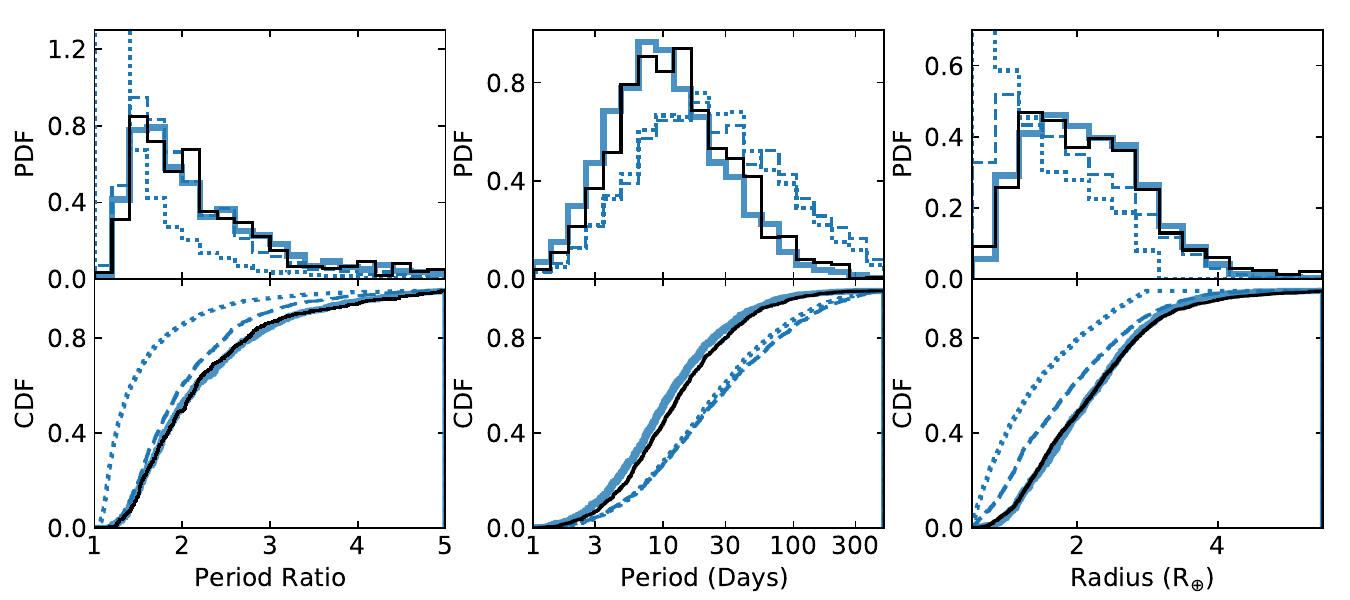}
\caption{PDF (top) and CDF (bottom) for adjacent-planet $\porb$ ratio (left), $\porb$ (middle), and $\Rp$ (right) for the fiducial ensemble \texttt{n8-e040-i024}. Line colors and styles denote the same populations as in \autoref{fig:K_dist_fidu}. When the dynamical state of the simulated population is similar to that observed, other observable properties show good agreement with the \kepler\ data (shown in black). The final radii of the synthetic planets from our model are obtained from the masses using the mass-radius relationship from \citet{Chen_2016}. 
}
\label{fig:fidu_dists}
\end{figure*}

\autoref{fig:K_dist_fidu} shows the distributions for $K$ from our fiducial ensemble. Clearly, $K \lesssim12$ is heavily sculpted by dynamical instabilities. The final model-detected $K$ distribution shows excellent agreement with that for the observed. A two-sample Kolmogorov–Smirnov (KS) test suggests that the null hypothesis that the observed and model-detected $K$ are drawn from the same underlying distribution can not be rejected with $p\gtrsim0.3$. This indicates that this ensemble of model planetary systems has achieved a similar dynamical state as \kepler's observed systems. 

Although, the initial distributions for $\porb$ ratios and $\Rp$ are significantly different from those observed, once a similar dynamical state is achieved, the key observable properties including the distribution of adjacent-planet $\porb$ ratios, $\porb$, and $\Rp$ show excellent agreement between the model-detected and observed populations (\autoref{fig:fidu_dists}). While the $\porb$ distribution remains relatively unchanged via dynamical sculpting, it is severely affected by \kepler's detection biases, making the $\porb$ distribution for the model-detected population resemble the \kepler\ data. A closer inspection shows that our model-detected $\porb$ distribution peaks at a marginally lower value compared to the observed (middle panel, \autoref{fig:fidu_dists}). This can be attributed to our initial conditions.

The simultaneous agreement in $K$ and $\porb$ ratios suggests that $\Mp$ in our models are consistent with the observed. This is also evident when we convert $\Mp$ into $\Rp$ and compare the model-detected population with the observed (right panel, \autoref{fig:fidu_dists}). The simulated systems initially only contained planets with $\Rp/R_\oplus<3$. Thus, all planets with $\Rp/R_\oplus\gtrsim3$ in our models are created by planet-planet collisions. Interestingly the model-detected $\Rp$ distribution shows excellent agreement with the \kepler\ data even for $\Rp/R_\oplus\gtrsim3$. This indicates the possibility that in nature, smaller planets form more abundantly, and most larger planets are formed via mergers of the smaller planets during dynamical instabilities.
On the other hand, we find a turnover in the final intrinsic PDF at $\Rp/R_\oplus\approx 1$, about $2\times$ higher than the smallest planet we have considered initially (\autoref{subsec:setup/star-planet-properties}). The turnover in the model-detected populations is at $\Rp/R_\oplus\gtrsim 1.5$, which illustrates \kepler's bias towards detecting larger planets. Since the turnovers in the final $\Rp$ distributions are at much higher values compared to the lower cutoff we used in our initial power-law $\Rp$ distribution, the effects of this initial choice are expected to be small.

It is also worth noting that in the sub-Neptune regime, $\Rp$ is strongly dependent on the atmosphere properties, and the ability for late-stage atmosphere retention by the planet can significantly affect the observed $\Rp$ distribution. Our idealised simulations do not incorporate additional physics for atmosphere loss, e.g., via collisions loss \citep[e.g.,][]{Hwang2017,Denman2020}, photoevaporation \citep[e.g.,][]{2017_Owen_Wu}, or core-powered evaporation \citep[e.g.,][]{2018_Ginzburg_CorePowered}. Hence, a comparison of the $\Rp$ distributions with more granular details \citep[e.g., the bimodal distribution of $\Rp$,][]{2017_Fulton} is beyond the scope of this work.

\begin{figure}
\includegraphics[width=\columnwidth]{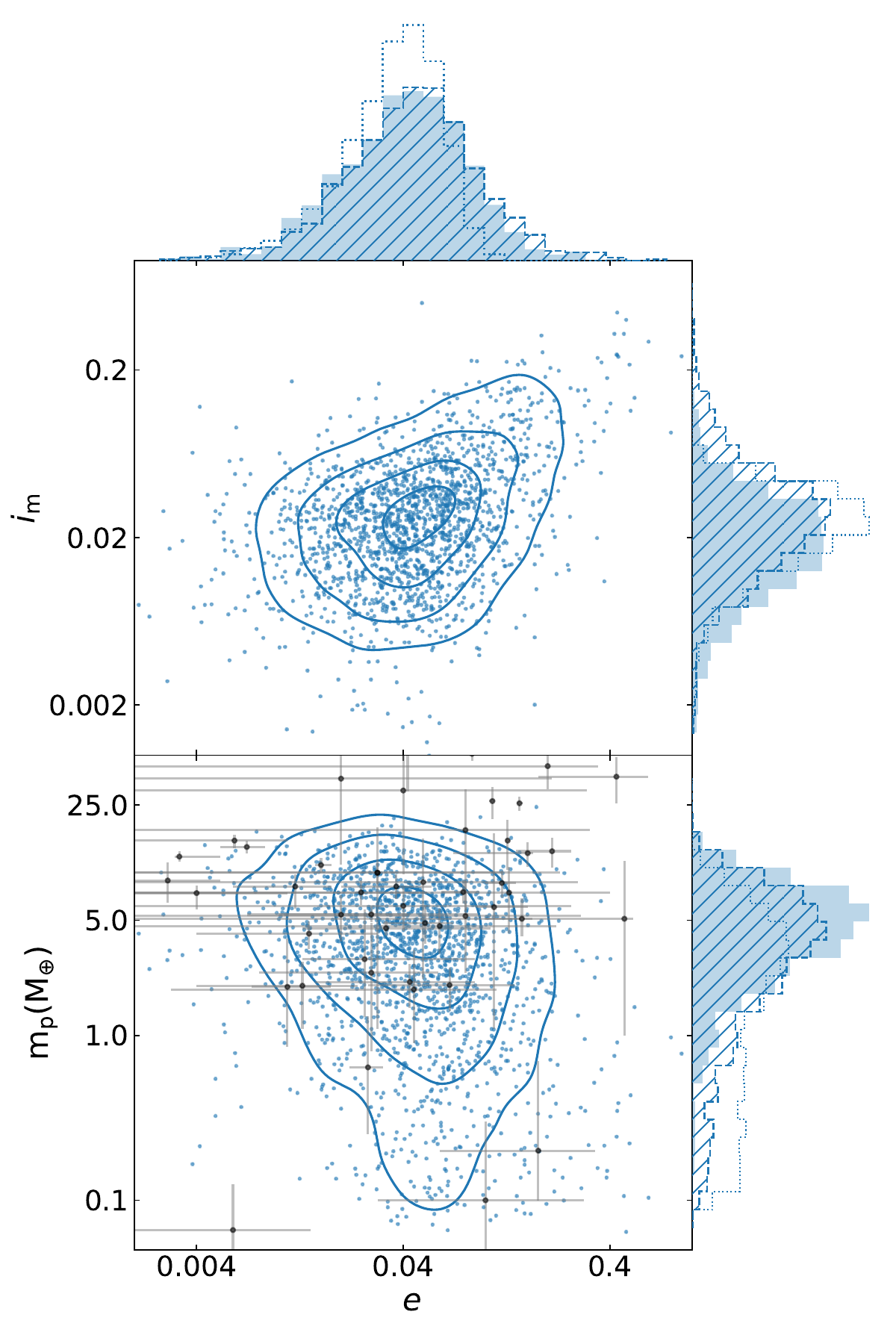}
\caption{$\ecc$ vs $\imut$ (top) and $\Mp$ (bottom) for the final intrinsic planetary systems in the fiducial ensemble. Dots denote individual planets and the contours confine $11.8\%$, $39.3\%$, $67.5\%$, and $86.5\%$ of all systems. The corresponding marginalised PDFs are shown on the top and side panels where dotted, dashed hatched, and filled histograms denote the initial, final intrinsic, and final model-detected populations, respectively. Black dots (with errorbars) denote \kepler\ planets with measured $\Mp$ and $\ecc$.
}
\label{fig:ecc_im_mp}
\end{figure}

We show the final $\ecc$, $\imut$, and $\Mp$ in \autoref{fig:ecc_im_mp}. The final $\ecc$ distribution of our model-detected population approximately follows a Rayleigh distribution with $\bar{\ecc}\approx0.042$ which agrees well with the population level observed $\ecc$ estimates for \kepler's multis \citep[][]{2016_Xie, 2019_Mills}. The model-detected single-planet systems exhibit a marginally higher $\bar{e}\approx0.048$ than the model-detected multis. However, the $\bar{\ecc}$ of the apparently single-planet systems are considerably lower compared to the estimated $\bar{e}\sim0.2-0.3$ for \kepler's singles \citep[e.g.,][]{2016_Xie, 2019_Mills, 2019_Van_Eylen} (more discussion on this later). 

In contrast to giant planet scattering \citep[e.g.,][]{Chatteerjee_2008}, $\imut$ does not increase significantly for most systems we have studied here. In our fiducial ensemble, $\imut$ increases from a Rayleigh distribution with $\bar{\imut}=0.024$ initially to $\bar{\imut}=0.026$ for the final intrinsic population. Nevertheless, the final intrinsic $\imut$ distribution exhibits a prominent tail extending to large $\imut$ (\autoref{fig:ecc_im_mp}).

The intrinsic final population shows a positive correlation between $\ecc$ and $\imut$ with a Pearson correlation coefficient, $\corr=0.56$, consistent with the estimate for the \kepler\ data from the `maximum AMD' model of \citet{2020_He}. We also find that $\Mp$ is anti-correlated with $\ecc$ with $\corr=-0.18$ (bottom panel, \autoref{fig:ecc_im_mp}). This is expected because the same level of AMD would alter the orbit of a lower-mass planet more compared to a higher-mass planet. Another subtle effect may also contribute to this anti-correlation. In our setup, large planets (e.g., $\Rp/R_\oplus>3$, equivalently $\Mp/M_\oplus\gtrsim 16$) are collision products. Since collisions reduce AMD (\autoref{fig:namd_evol}), the orbits of these planets are expected to be less eccentric.

\begin{figure}
\includegraphics[width=\columnwidth]{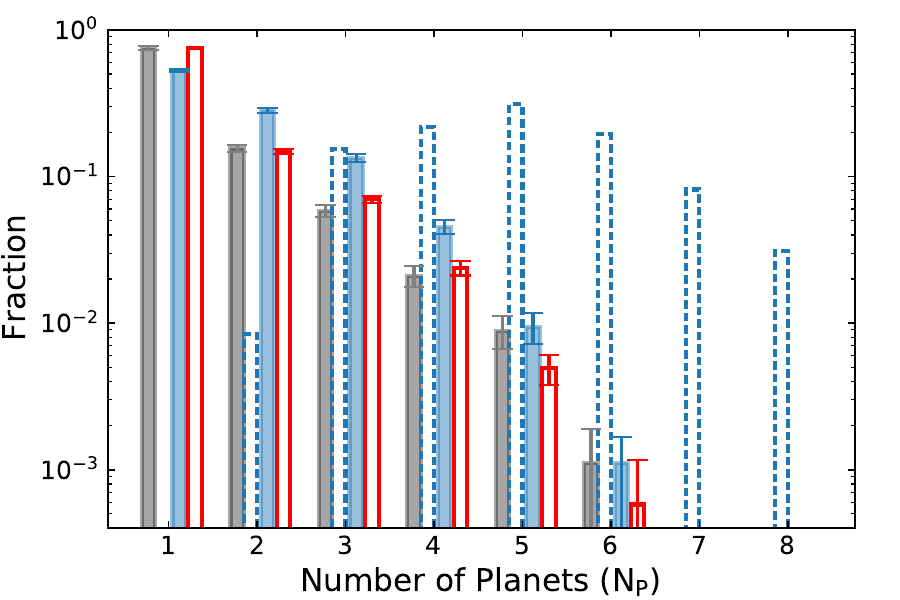}
\caption{Multiplicity distribution. Dashed blue, filled blue, and grey bars denote the final intrinsic, final model-detected, and observed populations, respectively. Red bars denote the multiplicity distribution after artificially adding single-planet systems ($f=0.48$) to model-detected population. Errorbars for model-detected denotes $1\sigma$ estimated using bootstrap with a sample size equal to the number of \kepler\ systems. Poisson errors are used for the \kepler\ data. There is an excess of observed single-transiting systems relative to our model-detected population. Injection of additional single-planet systems significantly improves the agreement.
}
\label{fig:multiplicity_dist}
\end{figure}

\autoref{fig:multiplicity_dist} shows the final multiplicity distributions of our simulated systems. Each planetary system in the fiducial ensemble has $\Npi=8$. Dynamical instabilities reduce $\Np$ predominantly due to collisions. Only $\sim3\%$ systems in this ensemble survive with all $8$ planets. The final intrinsic multiplicity ranges from 2 to 8. However, in most cases, only one planet transits. The detected number of systems declines steeply with increasing $\Np$. Although this trend is similar to the \kepler\ data, we find a lack of model-detected single-planet systems compared to what is observed. A similar discrepancy was noted in several past studies \citep[e.g.,][]{2011_Lissauer, Johansen_2012, 2013_Hansen_Murray, Ballard_2016} while attempting to explain the observed multiplicity distribution under various assumptions and is most commonly referred to as the ``\kepler\ dichotomy". The most common explanation of this apparent excess of single transiting systems is to consider a second population of either intrinsic single-planet systems or multis with higher $\imut$ \citep[e.g.,][]{Fang_2012, Mulders_2018, 2019_He}. Indeed, we find that the distribution of detected singles as well as the higher multiplicities can be explained simply by artificially injecting additional single transiting planets in the mix, such that the fraction of injected population, $f\sim 0.48$. 
This is only slightly higher than the estimated fraction of observed systems with high $\imut$ \citep[$f\sim0.4$;][]{Mulders_2018,2019_He}. Moreover, since our single transiting planets have lower $\ecc$ than the \kepler's singles, our results support the hypothesis that an additional population of dynamically hotter (or intrinsically single) planets may be needed to explain the apparent excess of observed singles. This may indicate a different dynamical origin for the dynamically hotter single-planet population or a population with higher-mass planets as perturbers (which were not included in this study). On the other hand, it has also been suggested that a carefully curated single-population model can also explain the \kepler\ data \citep[e.g.,][]{2018_Zhu, 2019_Sandford, 2020_He, Millholland_2021}.

\begin{figure}
\includegraphics[width=\columnwidth]{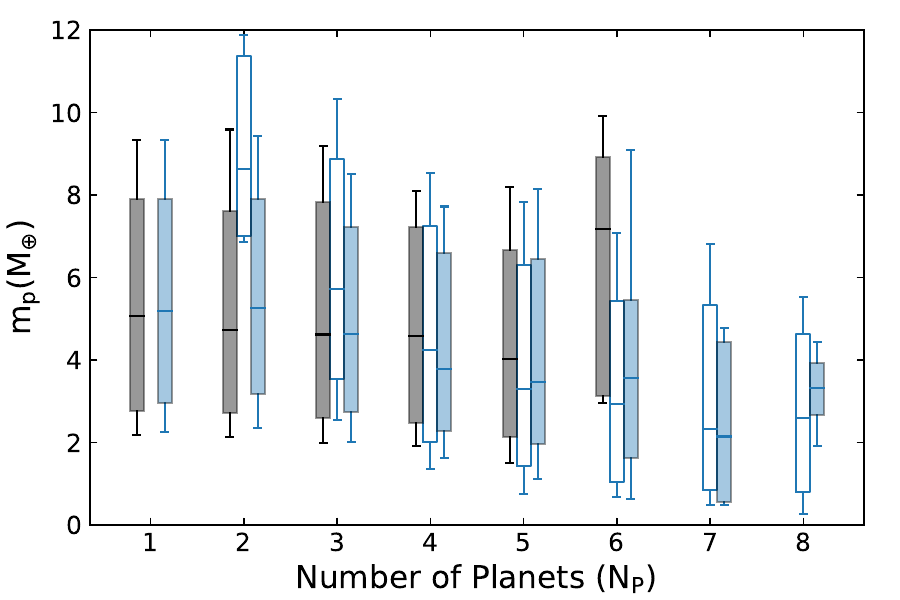}
\caption{Planetary multiplicity vs $\Mp$. The box and whiskers represent the $25$th--$75$th and $16$th-–$84$th percentiles. The unfilled and filled blue boxes denote the final intrinsic and model-detected populations. The filled grey boxes represent the \kepler\ planets. 
}
\label{fig:mass-npl}
\end{figure}
Interestingly, the low-multiplicity systems in our models are forged from systems with $\Npi=8$ predominantly via collisions. Thus, in the intrinsic population we find an anti-correlation between $\Mp$ and multiplicity. However, this trend is significantly washed-out in our model-detected population, which is in excellent agreement with the estimated $\Mp$ for the observed planets (\autoref{fig:mass-npl}). This trend in $\Mp$ with intrinsic multiplicities may provide a test for the dynamical models if these quantities are better understood for the observed planetary systems in the future. Of course, in reality, differences in the host mass and metallicity correlations (which we intentionally did not include) may make comparisons non-trivial.

\subsubsection{Peas In a Pod} \label{subsec:res/peas_in_a_pod}
One of the most interesting observed trends in \kepler's multis is that the planets within a particular system are more similar in size and regularly spaced compared to what would have been expected if they were drawn at random from all observed planets. This is commonly referred to as `peas in a pod' \citep{2018_Weiss}. Several studies argued that this trend likely was set early during the planet formation process \citep{Adams_2019, Adams_2020, Batygin_Morbi_2023}. As mentioned already, we do not introduce any intra-system uniformity in spacing or in planet properties. In fact, by design, the planets in our planetary systems are drawn completely randomly from the full set of possible planets and spacings. Hence, our simulations provide a clean test for the possibility of emergence of intra-system uniformity as a result of post-formation dynamical processes. Note that, since mass is more fundamental in our simulations, we quantify the intra-system uniformity in our models using the dimensionless mass uniformity metric 
\begin{equation}
\label{mass_unifomity_metric}
\mathcal{D} = \frac{1}{N_{\rm{sys}}} \sum_{j=1}^{N_{\rm{sys}}} \frac{\sigma_{\Mp_{j}}}{\bar{\Mp_{j}}},
\end{equation}
where $\sigma_{\Mp_{j}}$ and $\bar{\Mp_{j}}$ are the standard deviation and average $\Mp$ of the $j$th system, and $N_{\rm{sys}}$ is the number of systems in an ensemble \citep{Goldberg_2022}. Lower the value of $\mathcal{D}$, higher the intra-system uniformity. For \kepler's observed sub-jovian planetary systems $\mathcal{D}_{\rm{obs}}=0.49\pm0.01$.\footnote{$\Mp$ of the \kepler\ planets are estimated from the probabilistic $\Mp$--$\Rp$ relationship \citep{Chen_2016} the same way described in \autoref{subsec:setup/star-planet-properties}. The error bars for $\mathcal{D}_{\rm{obs}}$ are $1\sigma$ resulting from the intrinsic scatter in the $\Mp$--$\Rp$ relationship.} Initially, $\mathcal{D}\approx1$ in our ensemble, indicating complete lack of intra-system uniformity. We find that during dynamical instabilities, planet-planet collisions (and ejections in rare occassions) redistribute masses among the remaining planets in a way that decreases $\mathcal{D}$. For the final intrinsic and model-detected populations $\mathcal{D}=0.75$ and $0.49$, respectively. The latter is in excellent agreement with $\mathcal{D}_{\rm{obs}}$. This indicates that late-stage dynamical interactions in multi-planet systems can create the observed level of intra-system uniformity even if initially no such uniformity were present. Most recently, \citet[][, hereafter L23]{lammers2023intrasystem} come to the same conclusion independently despite a very different initial setup (see \autoref{appendix:l23} for a detailed discussion).

\begin{figure}
\includegraphics[width=\columnwidth]{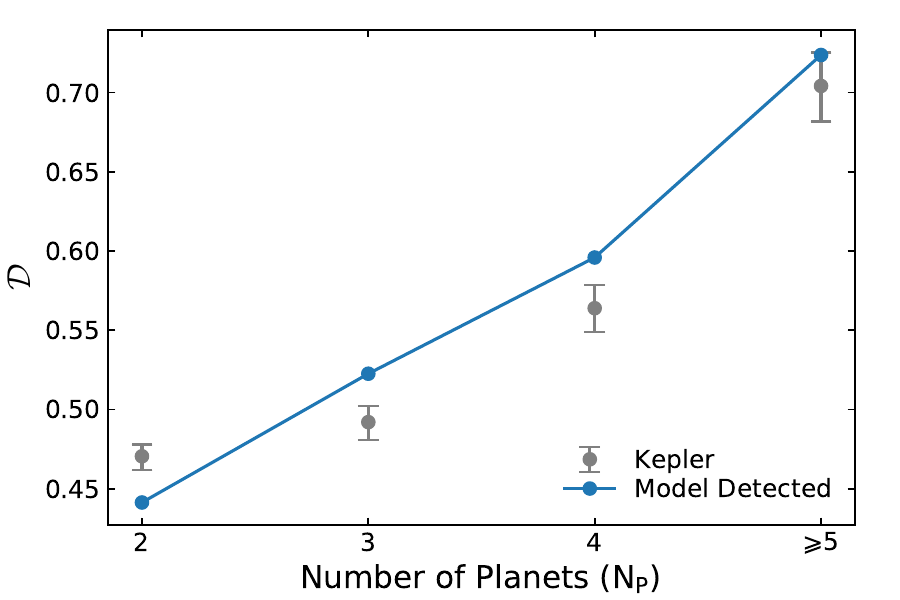}
\caption{$\mathcal{D}$ vs $\Np$ for our model-detected population (blue) and \kepler\ multis (grey). The error bars in $\mathcal{D}$ for the \kepler\ multis in each bin represent 25th and 75th percentiles resulting from the intrinsic scatter in the $\Mp$--$\Rp$ relationship from \citet{Chen_2016} within $1\sigma$.
}
\label{fig:uniformity_vs_multiplicity}
\end{figure}
Interestingly, \citet{Goldberg_2022} found that when started with a higher level of intra-system uniformity than observed, dynamical interactions, in fact, push systems towards reduced uniformity. On the other hand, L23 and this work independently come to the same conclusion that planetary systems with no initial intra-system uniformity, through dynamics, become more uniform. Thus, the observed level of uniformity in the \kepler's multis may represent a natural outcome of planetary dynamics independent of the initial setup. Investigating $\mathcal{D}_{\rm{obs}}$ as a function of observed $\Np$ may be a potential way to discern between whether the observed systems came from more or less initial intra-system uniformity. For example, if they were born more uniform and dynamical evolution reduced the uniformity to what we observe today, then the observed higher-multiplicity systems that are expected to have been less dynamically morphed should have a higher intra-system uniformity (lower $\mathcal{D}_{\rm obs}$) compared to the observed lower-multiplicity systems. On the other hand, if they were born with less intra-system uniformity and dynamics increased the uniformity, then the opposite trend would be observed. We find that $\mathcal{D}_{\rm{obs}}$ increases with increasing $\Np$ in the \kepler\ multis (\autoref{fig:uniformity_vs_multiplicity}). This trend favors the scenario where a less uniform initial population attains more intra-system uniformity through dynamics. Nevertheless, we caution the readers that the number of high-multiplicity observed systems decreases rapidly with increasing $\Np$, so a more careful investigation on this trend may be warranted in a future study. Also, note that the past studies do not take into account transit and detection biases like ours.

\begin{figure}
\includegraphics[width=\columnwidth]{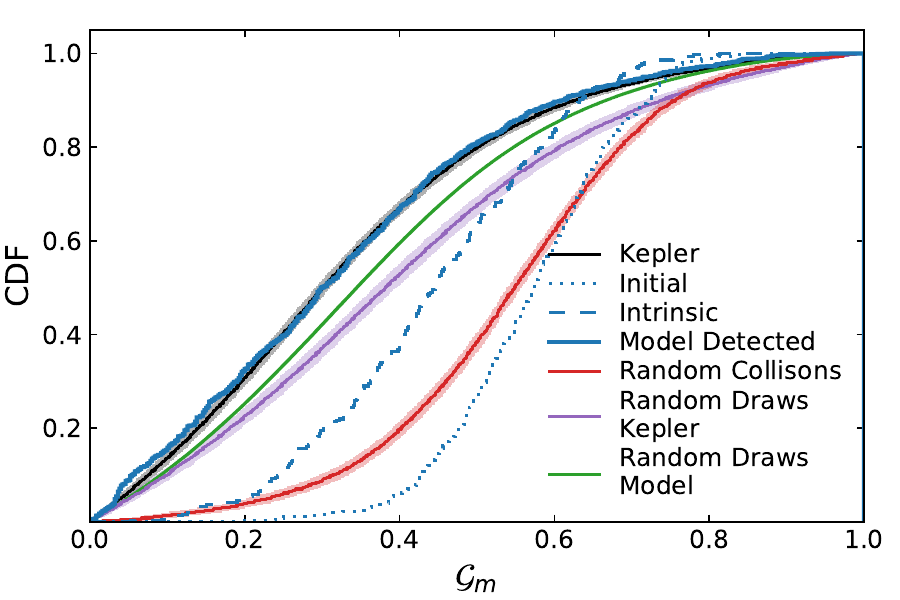}
\caption{CDF for the adjusted Gini index $\mathcal{G}_{m}$, indicating the distribution of intra-system uniformity in $\Mp$. The dotted, dashed, and thick solid blue lines denote the initial, the final intrinsic, and the model-detected populations. $\mathcal{G}_m$ for \kepler's multis is shown in black. Model-detected and observed intra-system uniformity for $\Mp$ is in agreement. The red line represents an artificial test population created by merging randomly chosen planet pairs from the initial planetary systems preserving the total number of collisions found in the $N$-body models. The purple (green) lines show the $\mathcal{G}_{m}$ distribution when the planets in each system are replaced by random planets from the entire \kepler\ (model-detected) population. Shades represent $1\sigma$.
}
\label{fig:m_gini_dist}
\end{figure}

While $\mathcal{D}$ shows the overall intra-system uniformity in a population of planetary systems, it does not show the distribution of uniformity in the systems within the population. We explore this using the adjusted Gini index \citep[$\mathcal{G}_{m}$;][]{Goyal_2022} for our ensemble.\footnote{$\mathcal{G}$ is routinely used for measuring the economic inequality in a population \citep{Gini_1912} which was later adjusted for small sample sizes by \citet{Deltas_2003}.} For a system with $\Np$ planets, we can write
\begin{equation}
\label{gini-index}
\mathcal{G}_{m} = \frac{1}{2\Np (\Np - 1) \bar{\Mp}} \sum_{j=1}^{\Np} \sum_{l=1}^{\Np}|\Mp_{j} - \Mp_{l}|.
\end{equation}
By construct, $\mathcal{G}_{m} = 0$ for a completely uniform system and tends to 1 for nonuniform systems. \autoref{fig:m_gini_dist} shows the distribution of $\mathcal{G}_{m}$ in our fiducial ensemble. Since we independently draw the initial planets in each system, initially $\mathcal{G}_{m}$ are very large (median $0.57$). Dynamical instabilities redistribute masses among the remaining planets to decrease $\mathcal{G}_{m}$ in each system. Thus, the $\mathcal{G}_m$ distribution shifts to lower values for the final intrinsic population indicating increased uniformity. Moreover, the $\mathcal{G}_m$ distribution for the final model-detected population is in excellent agreement with that for the observed multis. Thus, starting from planetary systems with completely uncorrelated planets, through a combination of dynamical encounters and transit and detection biases, our models produce planetary systems containing the same degree of intra-system uniformity as \kepler's multis.

We investigate this more in depth. To understand if the reduced $\mathcal{G}_m$ is due to the reduced $\Np$, we conduct a numerical experiment. We randomly choose pairs of planets from the initial planetary systems and merge them until the number of collisions is the same as that found in the $N$-body simulations. We find that the resulting planetary systems are not as uniform as the post-instability systems in our $N$-body simulations (\autoref{fig:m_gini_dist}). This shows that the level of intra-system uniformity in the final planetary systems is not simply due to a reduced $\Np$. The reason behind the induced uniformity is intricately related to the details of how planets of different $\Mp$ interact with each other as the NAMD of the system increases during dynamical instabilities. For example, orbits of lower-$\Mp$ planets would have higher excitation for the same NAMD which makes them more prone to collisions. Collisions between preferentially lower-$\Mp$ planets would tend to decrease the range in $\Mp$ in a system, thus increasing uniformity. 

\kepler's planets show similarities between planetary sizes within a system and are comparatively dissimilar to planets randomly drawn from other systems \citep{2018_Weiss}. This trend is apparent when the $\mathcal{G}_{m}$ distributions are compared for the observed population and a population created by replacing the planets in each multi with those drawn randomly from all observed multis (\autoref{fig:m_gini_dist}). Our model-detected population reproduce this trend qualitatively (\autoref{fig:m_gini_dist}). However, we find that our randomly drawn final detectable systems are somewhat less diverse than the observed inter-system diversity. This is expected. By design, we do not introduce any correlations between the host and the planets. In real systems, however, several birth and evolutionary conditions may introduce additional inter-system diversity relative to our controlled experiment.

Another aspect of `peas in a pod' is that the $\porb$ ratios between adjacent planet pairs within a system with $\Np \geq 3$ tend to be more correlated than those in the overall population of multis \citep{2018_Weiss}. We do not find any enhanced correlations arising from dynamics either in the intrinsic or the model-detected populations. If indeed the observed systems show significant additional intra-system regularities in spacings, it may have a different origin; for example, forming in resonant chains \citep{2017Izidoro, 2021_Izidoro, Goldberg_2022, 2023_Ghosh_Chatterjee} or a preferred initial $\porb$ distribution different from ours (L23).

\subsection{Variations In Initial Conditions} \label{subsec:res/IC_variations}

So far, we have discussed the effects of planetary dynamics using our fiducial ensemble (\texttt{n8-e040-i024}). Here, we examine whether our findings depend on the initial conditions by varying the initial properties including $\ecc$, $\inc$, and $\Npi$ one at a time. In each case, we compare the ensembles when the planetary systems achieve the same dynamical state as the observed (\autoref{subsec:setup/stopping}).

\begin{table*}
    \centering
    \caption{Median, 16 and 84 percentiles for final properties for the intrinsic and model-detected populations. $\bar{e}$ and $\bar{\imut}$ denotes the mean of the best-fit Rayleigh distributions for eccentricities and mutual inclinations (measured with respect to the system invariant plane).}
    \label{tab:results_table}
        \setlength{\tabcolsep}{5pt}
        \setlength{\extrarowheight}{4pt}
        \begin{threeparttable}
    \begin{tabular}{c @{\hskip 0.1in}@{\vline}@{\hskip 0.1in} cccccc @{\hskip 0.1in}@{\vline}@{\hskip 0.1in} cccccc}
        \hline
            \hline
            \multicolumn{1}{c @{\hskip 0.1in}@{\vline}@{\hskip 0.1in}}{Ensemble} & \multicolumn{6}{c @{\hskip 0.1in}@{\vline}@{\hskip 0.1in}}{Intrinsic} & \multicolumn{5}{c}{Detected}\\
        Name & $K$ & $\porb$ ratio& $\porb$/day & $\Mp/M_{\earth}$  & $\bar{\ecc}$ & $\bar{\imut}$ & $K$ & $\porb$ ratio& $\porb$/day & $\Mp/M_{\earth}$  & $\bar{\ecc}$ \\
        \hline
              n8-e040-i024* &  $20.7^{10.4}_{-7.7}$ &  $1.85^{0.76}_{-0.39}$ &   $23.6^{88.9}_{-17.0}$ &  $3.4^{4.5}_{-2.6}$ &  $0.047$ &  $0.026$ &  $21.0^{14.1}_{-8.6}$ &  $2.03^{1.25}_{-0.51}$ &  $9.2^{17.8}_{-5.6}$ &  $5.0^{4.0}_{-2.8}$ &  $0.044$ \\
              n8-e025-i024 &  $20.6^{10.2}_{-7.5}$ &  $1.86^{0.76}_{-0.39}$ &   $22.9^{82.1}_{-16.1}$ &  $3.6^{4.4}_{-2.7}$ &  $0.036$ &  $0.025$ &  $21.0^{13.7}_{-8.6}$ &  $1.98^{1.29}_{-0.47}$ &  $9.2^{18.4}_{-5.7}$ &  $5.0^{4.1}_{-2.8}$ &  $0.035$ \\
              n8-e030-i024 &  $20.8^{10.0}_{-7.4}$ &   $1.88^{0.75}_{-0.4}$ &   $22.9^{84.3}_{-16.2}$ &  $3.6^{4.4}_{-2.7}$ &  $0.039$ &  $0.025$ &  $21.0^{14.1}_{-8.3}$ &  $2.02^{1.25}_{-0.51}$ &  $9.3^{18.7}_{-5.8}$ &  $5.0^{3.9}_{-2.8}$ &  $0.037$ \\
              n8-e035-i024 &  $20.8^{10.6}_{-7.4}$ &   $1.89^{0.79}_{-0.4}$ &   $22.6^{84.7}_{-16.1}$ &  $3.6^{4.3}_{-2.7}$ &  $0.041$ &  $0.026$ &  $21.2^{14.2}_{-8.3}$ &  $2.04^{1.25}_{-0.51}$ &  $9.2^{19.5}_{-5.6}$ &  $5.1^{3.8}_{-2.9}$ &  $0.039$ \\
              n8-e050-i024 &  $21.1^{10.3}_{-7.8}$ &  $1.91^{0.77}_{-0.42}$ &   $23.2^{91.3}_{-16.4}$ &  $3.6^{4.5}_{-2.7}$ &  $0.053$ &  $0.026$ &  $21.2^{14.3}_{-8.2}$ &  $2.03^{1.37}_{-0.51}$ &  $9.3^{18.1}_{-5.7}$ &  $4.9^{4.2}_{-2.7}$ &   $0.05$ \\
              n8-e060-i024 &  $19.9^{10.1}_{-7.6}$ &  $1.79^{0.74}_{-0.39}$ &  $25.8^{102.1}_{-18.8}$ &  $3.1^{4.5}_{-2.5}$ &  $0.064$ &  $0.027$ &  $21.4^{13.4}_{-8.8}$ &  $2.04^{1.23}_{-0.51}$ &  $9.1^{16.7}_{-5.6}$ &  $4.8^{4.2}_{-2.7}$ &  $0.057$ \\
              n8-e080-i024 &  $17.8^{11.0}_{-8.3}$ &  $1.63^{0.69}_{-0.37}$ &   $27.6^{85.3}_{-20.4}$ &  $2.6^{4.5}_{-2.3}$ &  $0.083$ &  $0.027$ &  $21.4^{13.9}_{-9.1}$ &  $2.04^{1.31}_{-0.53}$ &  $8.9^{18.2}_{-5.5}$ &  $4.7^{4.1}_{-2.8}$ &  $0.069$ \\
              n8-e040-i010 &  $21.3^{10.2}_{-8.0}$ &  $1.92^{0.76}_{-0.42}$ &   $22.3^{85.2}_{-15.8}$ &  $3.8^{4.3}_{-2.8}$ &  $0.044$ &  $0.015$ &  $20.8^{12.2}_{-8.4}$ &  $2.04^{1.11}_{-0.53}$ &  $9.3^{19.3}_{-5.6}$ &  $5.1^{4.0}_{-2.8}$ &  $0.043$ \\
              n8-e040-i050 &   $19.8^{9.9}_{-7.7}$ &  $1.76^{0.73}_{-0.37}$ &   $25.5^{99.5}_{-18.6}$ &  $3.1^{4.4}_{-2.5}$ &  $0.049$ &  $0.048$ &  $21.6^{14.7}_{-8.9}$ &  $2.08^{1.53}_{-0.56}$ &  $8.7^{18.1}_{-5.4}$ &  $4.9^{4.1}_{-2.8}$ &  $0.045$ \\
              n8-e040-i075 &  $19.3^{10.1}_{-7.9}$ &  $1.73^{0.66}_{-0.37}$ &   $25.7^{97.5}_{-18.6}$ &  $2.9^{4.6}_{-2.5}$ &  $0.054$ &  $0.068$ &  $21.7^{16.2}_{-9.0}$ &  $2.14^{1.64}_{-0.62}$ &  $8.7^{17.4}_{-5.4}$ &  $4.7^{4.2}_{-2.7}$ &   $0.05$ \\
              n6-e040-i024 &  $20.7^{10.5}_{-8.0}$ &  $1.86^{0.72}_{-0.41}$ &   $15.2^{41.4}_{-10.2}$ &  $3.4^{4.1}_{-2.6}$ &  $0.042$ &  $0.022$ &  $19.9^{12.3}_{-8.3}$ &  $1.89^{1.03}_{-0.43}$ &  $8.3^{14.9}_{-4.9}$ &  $4.5^{4.2}_{-2.6}$ &  $0.039$ \\
             n10-e040-i024 &  $20.5^{10.6}_{-7.7}$ &  $1.84^{0.82}_{-0.41}$ &  $40.0^{227.5}_{-32.3}$ &  $3.5^{4.2}_{-2.8}$ &  $0.052$ &  $0.031$ &  $21.3^{13.7}_{-8.4}$ &   $2.05^{1.5}_{-0.52}$ &  $9.0^{21.0}_{-5.7}$ &  $5.0^{3.8}_{-2.6}$ &  $0.045$ \\
             \kepler & - & - & - & - & - & - & $22.0^{16.2}_{-8.5}$ & $2.09^{1.65}_{-0.56}$ & $11.4^{28.3}_{-7.6}$ & $4.9^{4.4}_{-2.8}$ & $0.044 \pm 0.015$ \tnote{a}&\\
        \hline
    \end{tabular}
         \begin{tablenotes}
            \item[a] For \kepler's multi-transiting systems \citep{2019_Mills}.
        \end{tablenotes}
     \end{threeparttable}
\end{table*}
\begin{table*}
    \centering
    \caption{Fractions ($f_{\Np,j}$) of planetary systems with $\Np=j$. Errors represent $1\sigma$ estimated using bootstrap with a sample size equal to the number of \kepler\ systems. The limits in the \kepler\ dataset represents Poisson errors. }
    \label{tab:multiplicity}
        \setlength{\tabcolsep}{1pt}
        \setlength{\extrarowheight}{4pt}
        \begin{threeparttable}
    \begin{tabular}{c @{\hskip 0.1in}@{\vline}@{\hskip 0.1in} ccccccc @{\hskip 0.1in}@{\vline}@{\hskip 0.1in} cccccc}
 
        \hline
            \hline
            \multicolumn{1}{c @{\hskip 0.1in}@{\vline}@{\hskip 0.1in}}{Ensemble} & \multicolumn{7}{c @{\hskip 0.1in}@{\vline}@{\hskip 0.1in}}{Intrinsic} & \multicolumn{6}{c}{Detected}\\
        Name & $f_{\Np,2}$ & $f_{\Np,3}$ & $f_{\Np,4}$ &$f_{\Np,5}$  & $f_{\Np,6}$ & $f_{\Np,7}$ & $f_{\Np,8}$ & $f_{\Np,1}$ & $f_{\Np,2}$ & $f_{\Np,3}$  & $f_{\Np,4}$ & $f_{\Np,5}$ & $f_{\Np,>=6}$ \\
            \hline
                n8-e040-i024*  &  $0.01$ &  $0.15$ &  $0.22$ &  $0.31$ &  $0.19$ &  $0.08$ &  $0.03$ &  $0.53^{0.012}_{-0.012}$ &  $0.28^{0.011}_{-0.011}$ &  $0.13^{0.008}_{-0.008}$ &    $0.046^{0.005}_{-0.005}$ &  $0.009^{0.0022}_{-0.0022}$ &  $0.001^{0.0011}_{-0.0006}$ \\
                n8-e025-i024  &  $0.02$ &  $0.12$ &  $0.27$ &   $0.3$ &  $0.18$ &  $0.08$ &  $0.03$ &  $0.52^{0.012}_{-0.011}$ &   $0.29^{0.011}_{-0.01}$ &  $0.14^{0.008}_{-0.008}$ &  $0.045^{0.0049}_{-0.0044}$ &   $0.01^{0.0027}_{-0.0022}$ &  $0.001^{0.0005}_{-0.0005}$ \\
                n8-e030-i024  &  $0.02$ &  $0.14$ &  $0.28$ &  $0.27$ &  $0.18$ &  $0.08$ &  $0.03$ &  $0.53^{0.012}_{-0.011}$ &   $0.28^{0.011}_{-0.01}$ &  $0.14^{0.008}_{-0.008}$ &  $0.045^{0.0049}_{-0.0049}$ &  $0.009^{0.0027}_{-0.0022}$ &  $0.001^{0.0005}_{-0.0011}$ \\
                n8-e035-i024  &  $0.01$ &  $0.12$ &  $0.32$ &  $0.29$ &  $0.18$ &  $0.06$ &  $0.02$ &  $0.53^{0.011}_{-0.011}$ &    $0.29^{0.01}_{-0.01}$ &  $0.13^{0.008}_{-0.008}$ &  $0.044^{0.0049}_{-0.0049}$ &   $0.01^{0.0022}_{-0.0027}$ &  $0.001^{0.0011}_{-0.0005}$ \\
                n8-e050-i024  &  $0.02$ &  $0.15$ &  $0.27$ &  $0.27$ &   $0.2$ &  $0.07$ &  $0.02$ &  $0.53^{0.012}_{-0.012}$ &   $0.29^{0.011}_{-0.01}$ &  $0.13^{0.008}_{-0.008}$ &  $0.042^{0.0049}_{-0.0044}$ &  $0.009^{0.0022}_{-0.0022}$ &  $0.001^{0.0005}_{-0.0005}$ \\
                n8-e060-i024  &  $0.01$ &  $0.09$ &  $0.24$ &  $0.22$ &  $0.23$ &  $0.15$ &  $0.06$ &  $0.52^{0.012}_{-0.012}$ &  $0.28^{0.011}_{-0.011}$ &  $0.14^{0.009}_{-0.008}$ &  $0.043^{0.0051}_{-0.0045}$ &  $0.009^{0.0023}_{-0.0023}$ &  $0.001^{0.0011}_{-0.0006}$ \\
                n8-e080-i024  &   $0.0$ &  $0.05$ &  $0.12$ &  $0.19$ &  $0.27$ &  $0.28$ &  $0.09$ &  $0.52^{0.013}_{-0.012}$ &   $0.28^{0.011}_{-0.01}$ &  $0.14^{0.008}_{-0.008}$ &  $0.048^{0.0048}_{-0.0054}$ &   $0.01^{0.0024}_{-0.0024}$ &  $0.001^{0.0012}_{-0.0006}$ \\
                n8-e040-i010  &  $0.01$ &  $0.17$ &  $0.31$ &  $0.28$ &  $0.15$ &  $0.05$ &  $0.03$ &   $0.5^{0.012}_{-0.011}$ &   $0.29^{0.01}_{-0.011}$ &  $0.15^{0.008}_{-0.008}$ &  $0.044^{0.0049}_{-0.0043}$ &  $0.011^{0.0022}_{-0.0022}$ &  $0.002^{0.0011}_{-0.0005}$ \\
                n8-e040-i050  &  $0.01$ &  $0.09$ &  $0.19$ &  $0.25$ &  $0.26$ &  $0.15$ &  $0.05$ &  $0.56^{0.012}_{-0.012}$ &  $0.28^{0.011}_{-0.011}$ &  $0.12^{0.008}_{-0.007}$ &  $0.033^{0.0047}_{-0.0041}$ &  $0.006^{0.0018}_{-0.0018}$ &  $0.001^{0.0012}_{-0.0006}$ \\
                n8-e040-i075  &  $0.02$ &  $0.05$ &   $0.2$ &  $0.22$ &  $0.27$ &  $0.17$ &  $0.08$ &   $0.6^{0.012}_{-0.012}$ &   $0.27^{0.011}_{-0.01}$ &   $0.1^{0.008}_{-0.007}$ &  $0.023^{0.0037}_{-0.0037}$ &  $0.003^{0.0012}_{-0.0012}$ &        $0.0^{0.0006}_{0.0}$ \\
                n6-e040-i024  &  $0.07$ &  $0.34$ &  $0.39$ &  $0.18$ &  $0.02$ &       - &       - &  $0.54^{0.011}_{-0.012}$ &   $0.3^{0.011}_{-0.011}$ &  $0.12^{0.008}_{-0.008}$ &  $0.028^{0.0044}_{-0.0039}$ &  $0.004^{0.0011}_{-0.0017}$ &           $0.0^{0.0}_{0.0}$ \\
                n10-e040-i024 &       - &  $0.01$ &  $0.09$ &  $0.19$ &  $0.28$ &  $0.18$ &  $0.25$ \tnote{a}&  $0.54^{0.012}_{-0.012}$ &  $0.27^{0.011}_{-0.011}$ &  $0.12^{0.008}_{-0.008}$ &  $0.049^{0.0053}_{-0.0053}$ &    $0.016^{0.003}_{-0.003}$ &  $0.004^{0.0018}_{-0.0012}$ \\
                Kepler        &       - &       - &       - &       - &       - &       - &       - &            $0.76\pm0.02$ &           $0.16\pm0.009$ &           $0.06\pm0.006$ &            $0.021\pm0.0034$ &            $0.009\pm0.0022$ &            $0.001\pm0.0008$ \\
        \hline
    \end{tabular}
        \begin{tablenotes}
            \item[a] This ensemble contains intrinsic planetary systems with $\Np=9$ ($f_{\Np,9}=0.05$) and $10$ ($f_{\Np,10}=0.03$). We combine them with $f_{\Np,8}$.
        \end{tablenotes}
        \end{threeparttable}
\end{table*}
\subsubsection{Effects of the initial eccentricity distribution} \label{subsec:res/ecc-effect}
\begin{figure}
\includegraphics[width=\columnwidth]{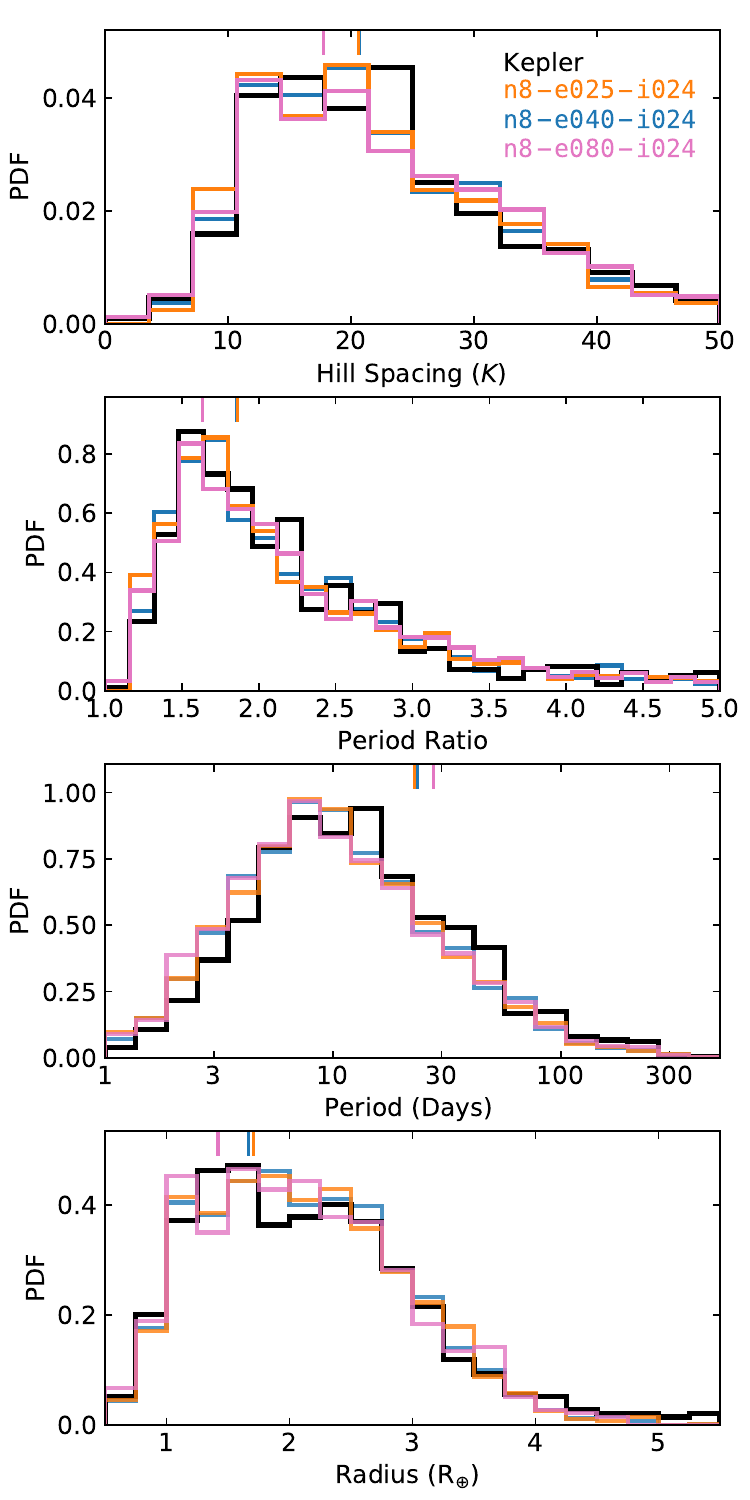}
\caption{PDF of $K$, adjacent-planet $\porb$ ratios, $\porb$, and $\Rp$ (top to bottom) for the model-detected populations from ensembles \texttt{n8-e025-i024} (orange), \texttt{n8-e040-i024}(blue), and \texttt{n8-e080-i024} (pink). Vertical lines near the top indicate the medians for the intrinsic distributions in these ensembles. \kepler\ data is shown in black.
}
\label{fig:dists-eccs}
\end{figure}
\begin{figure}
\includegraphics[width=\columnwidth]{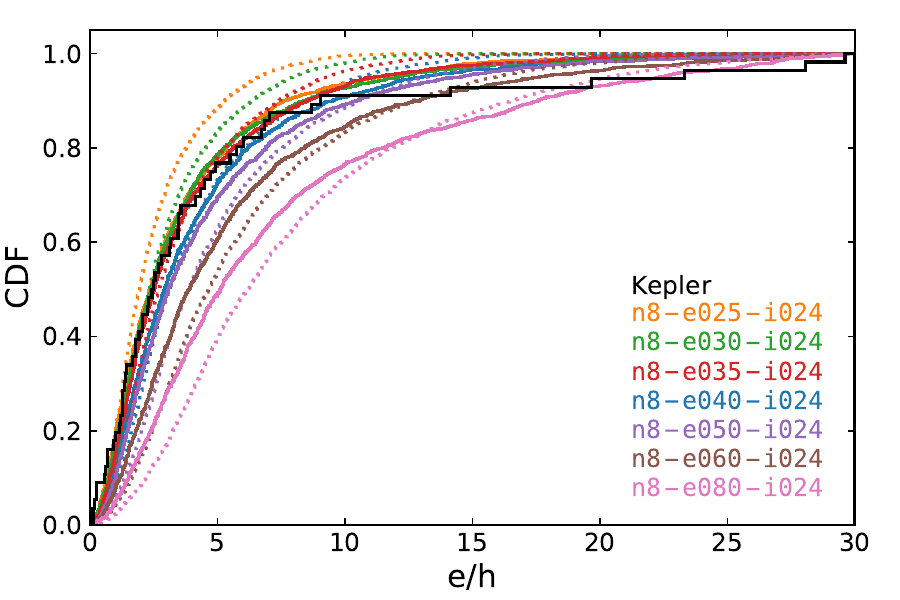}
\caption{CDF for $\ecc/h$ for ensembles created with different initial $\bar{e}$. Solid (dotted) lines denote the final intrinsic (initial) population. The black line represents \kepler's planets with measured $\Mp$ and $\ecc$. 
}
\label{fig:ecc-dist}
\end{figure}

Higher initial $\bar{\ecc}$ leads to earlier onset of instabilities. Hence, ensembles with higher initial $\bar{\ecc}$ reach the desired dynamical state earlier (\autoref{tab:initial_props}). Nevertheless, when an ensemble reaches the same dynamical state as observed, independent of the initial $\bar{\ecc}$, the distributions of observable properties are consistent with those for the fiducial ensemble and the observed (\autoref{fig:dists-eccs}). Note that, since the desired dynamical state is determined by comparing the model-detected population with the observed, and the detectability of a particular planet does depend on $\ecc$ via differences in the transit duration, small differences may remain in the final intrinsic distributions.

\autoref{fig:ecc-dist} shows the initial and final $\ecc$ distributions scaled by the reduced Hill radius $h\equiv(\frac{\Mp}{3 M_{*}})^{1/3}$, for several ensembles differing from each other only in the initial $\bar{\ecc}$. We find that for $\bar{e}\lesssim0.035$, the final intrinsic $e/h$ distributions are approximately similar to each other. Moreover, the $e/h$ distributions move towards higher values through dynamics in these ensembles with low initial $\bar{\ecc}$. In contrast, for ensembles with initial $\bar{e}\gtrsim0.04$, the final intrinsic $e/h$ distributions move towards lower values compared to the initial. We find that the observed $e/h$ distribution for \kepler's multis is consistent with our ensembles with lower initial eccentricities. Interestingly, because of the contrasting nature of how $e/h$ distributions evolve via dynamical interactions in initially low and high $\bar{\ecc}$ ensembles, it appears that independent of the initial eccentricities, the final $e/h$ distribution always approaches the observed $e/h$ distribution. Nevertheless, note that at present it is challenging to undertake an unbiased detailed study of this given the very small ($56$) number of \kepler\ planets with both $\Mp$ and $\ecc$ measurements.

\begin{figure}
\includegraphics[width=\columnwidth]{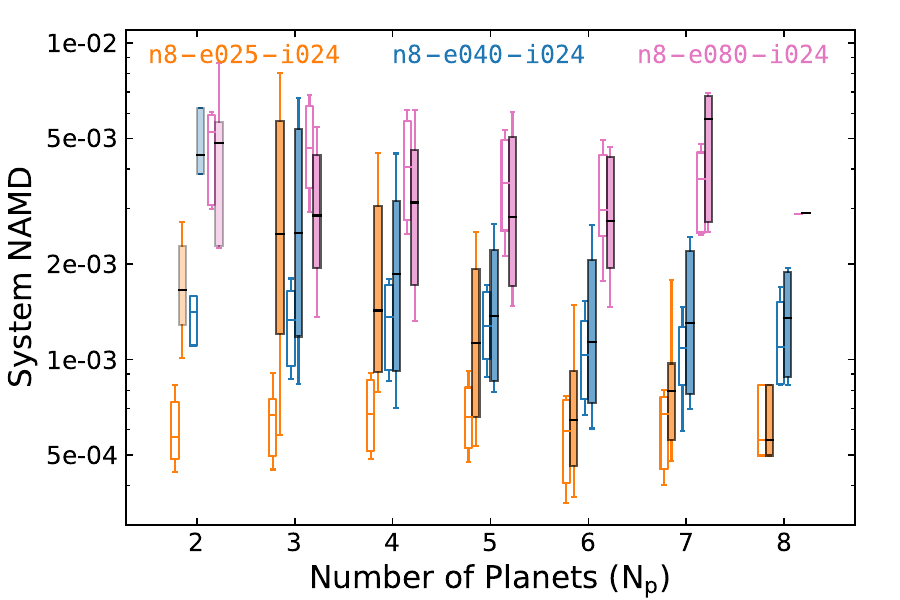}
\caption{$\Np$ vs NAMD for the final intrinsic populations in ensembles \texttt{n8-e025-i024} (orange), \texttt{n8-e040-i024} (blue) and \texttt{n8-e080-i024} (pink). The box and whiskers represent the $25$th–$75$th and $16$th–$84$th percentiles. The filled boxes represent the final intrinsic NAMD distributions in each bin, and the empty boxes represent the initial NAMD distributions of the same systems. Note that the NAMD values for the bin with $\Np=2$ may be unreliable due to very few data points (\autoref{tab:multiplicity}).
}
\label{fig:namd_npl}
\end{figure}

We find an anti-correlation between the NAMD of the planetary systems and multiplicity, except for ensembles with very high initial NAMD (\autoref{fig:namd_npl}). The anti-correlation in our models is easy to understand. Planet-planet interactions increase the overall NAMD in a system, while collisions and ejections decrease it. However, the decrease is usually not sufficient to bring the NAMD back to the pre-scattering level (\autoref{fig:namd_evol}). In our models, the intrinsically lower-multiplicity systems have gone through more dynamical activities, hence, they tend to exhibit higher NAMD compared to high-multiplicity systems. A similar anti-correlation has been inferred for the observed multis \citep{2020_He,2020_Turrini}. This trend, however, disappears in ensembles with very high initial $\bar{e}$ ($\gtrsim0.04$). In those ensembles, the initial NAMD is so high that even the high-multiplicity systems that are less dynamically evolved, retain the initial high NAMD. So our high eccentricity ensembles, or in general, the ensembles with high initial NAMD, are not commensurate with the observed data if the inferred NAMD anti-correlation with multiplicity is real. This likely indicates that the pre-instability Kepler planets had minimal orbital excitation. On the other hand, the NAMD values for the observed high-multiplicity systems may be effectively used to put constraints on the initial NAMD since there is not much difference between the final model detected NAMD and initial.

As expected, this trend is directly translated to orbital eccentricities. The anti-correlation between $\ecc$ and $\Np$ in the final intrinsic population ranges from $\corr=-0.18$ in the lowest initial $\bar{\ecc}$ ensemble (\texttt{n8-e025-i024}) to $\corr=0.03$ in the highest (\texttt{n8-e080-i024}). The inferred anti-correlation for the observed multis is qualitatively similar to those in the ensembles with low initial $\bar{\ecc}$ ensembles \citep{2015_Limbach,2017_Zinzi,2020_Bach,2020_He}. 

Other properties such as the multiplicity distributions of the model-detected populations do not show any significant differences depending on the initial $\bar{\ecc}$ (\autoref{tab:results_table}, \autoref{tab:multiplicity}).

\subsubsection{Effect of Initial Inclination} \label{subsec:res/inc-effect}

\begin{figure}
\includegraphics[width=\columnwidth]{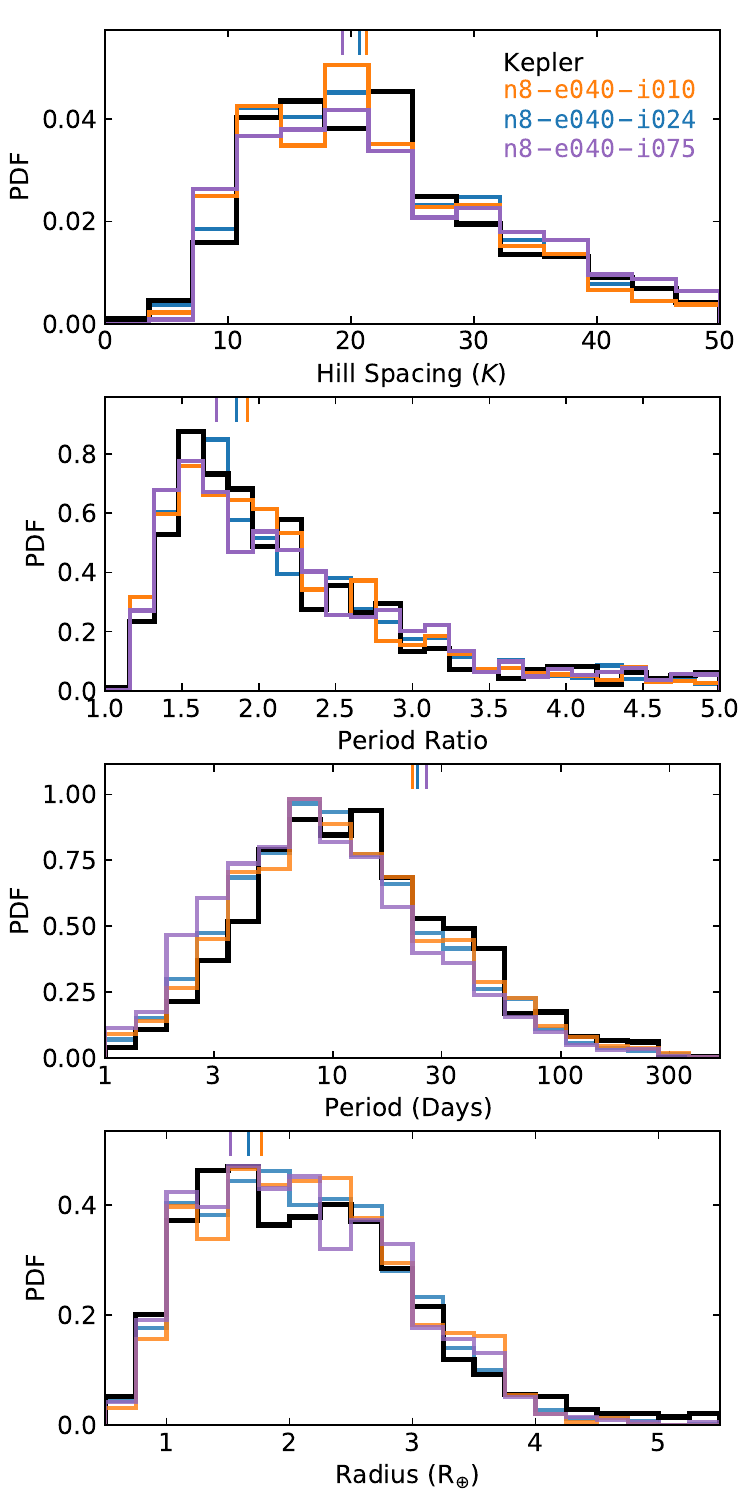}
\caption{Similar to \autoref{fig:dists-eccs} but for ensembles \texttt{n8-e040-i010} (orange), \texttt{n8-e040-i024} (blue) and \texttt{n8-e040-i075} (purple).}
\label{fig:dists-incs}
\end{figure}

Similar to the finding in the case of ensembles with different initial $\bar{\ecc}$, we find that when the dynamical states of our ensembles are consistent with that of \kepler's multis, the distributions of other observable properties are consistent with each other and the observed, independent of the initial $\bar{\inc}$ (\autoref{fig:dists-incs}). The only difference is that the ensembles with higher initial $\bar{\inc}$ reach the target dynamical state earlier (\autoref{tab:initial_props}). Of course, $\imut$ plays a significant role in determining transit probability in multis. As a result, when the $K$ distributions for the model-detected populations are consistent with the observed, the intrinsic distributions may show subtle differences. For example, the intrinsic distributions for $K$, $\porb$ ratios, and $\Rp$ shift to lower values with increasing initial $\bar{\inc}$ (\autoref{tab:results_table}).

\begin{figure}
\includegraphics[width=\columnwidth]{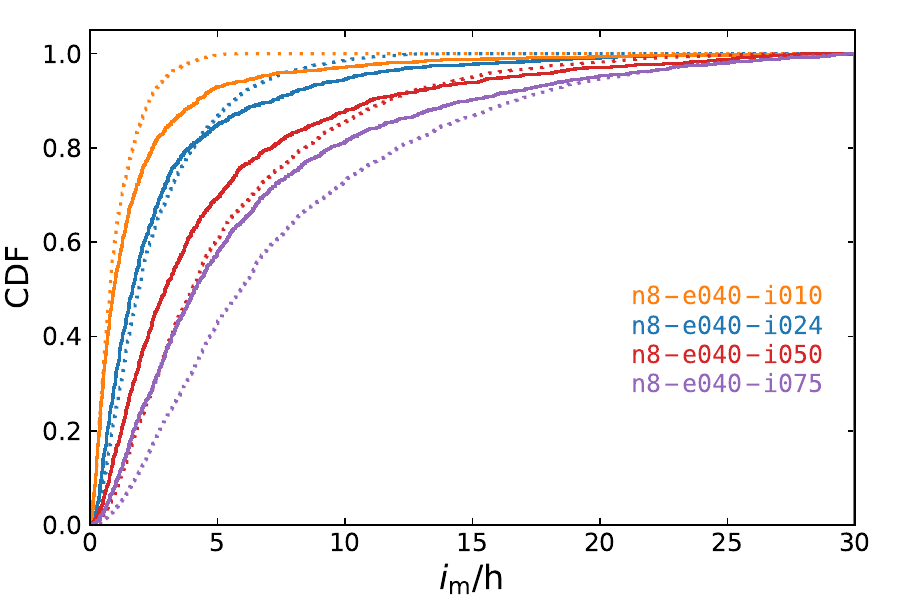}
\caption{CDF for $\imut/h$ for ensembles (see legend) with different initial $\bar{\inc}$. Solid (dotted) lines denote the final intrinsic (initial) populations. 
}
\label{fig:inc-dist}
\end{figure}

The final intrinsic $\imut/h$ distribution depends on the initial $\bar{\inc}$ (\autoref{fig:inc-dist}). Through dynamical evolution the $\imut/h$ distribution shifts to lower (higher) values in the ensembles created with initial $\bar{\inc}=0.05$ ($0.01$). For the ensemble with a medium initial $\bar{\inc}=0.024$, the $\imut/h$ distribution shows little change. Thus, it appears that a natural outcome of dynamical evolution (for the class of multis studied here) is a preferred intrinsic $\imut/h$ distribution; planetary systems initially more aligned than this become more misaligned and vice-versa. This trend is qualitatively similar to the trend with $e/h$ distributions for ensembles modeled with different initial $\bar{\ecc}$ (\autoref{subsec:res/ecc-effect}). It will be very interesting to verify whether indeed the intrinsic $\imut$ and $\ecc$ distributions exhibit such a tendency. However, estimating intrinsic $\imut$ and $\ecc$ distribution for the observed multis, ideally in a model agnostic way, is challenging and is beyond the scope of this work.

We find some subtle effects in the multiplicity distribution depending on the initial $\bar{\inc}$ (\autoref{tab:multiplicity}). For example, ensemble \texttt{n8-e040-i050} with initial $\bar{\inc}=0.05$ intrinsically retains more systems with $\Np>6$. However, due to higher final $\imut$, these planets have lower transit probabilities. As a result, the fraction of detected systems with $\Np>4$ is lower in this ensemble relative to the ensemble with $\bar{\inc}=0.01$ (\texttt{n8-e040-i010}, \autoref{tab:multiplicity}). While the final intrinsic $\imut$ distribution has large spreads in all ensembles, the median $\imut$ shows a modest anti-correlation with $\Np$, especially in the initially low $\bar{\inc}$ ensemble ($\corr=-0.11$ for \texttt{n8-e040-i010}).

\subsubsection{Effect of Initial Number of Planets} \label{subsec:res/np-effect}
\begin{figure}
\includegraphics[width=\columnwidth]{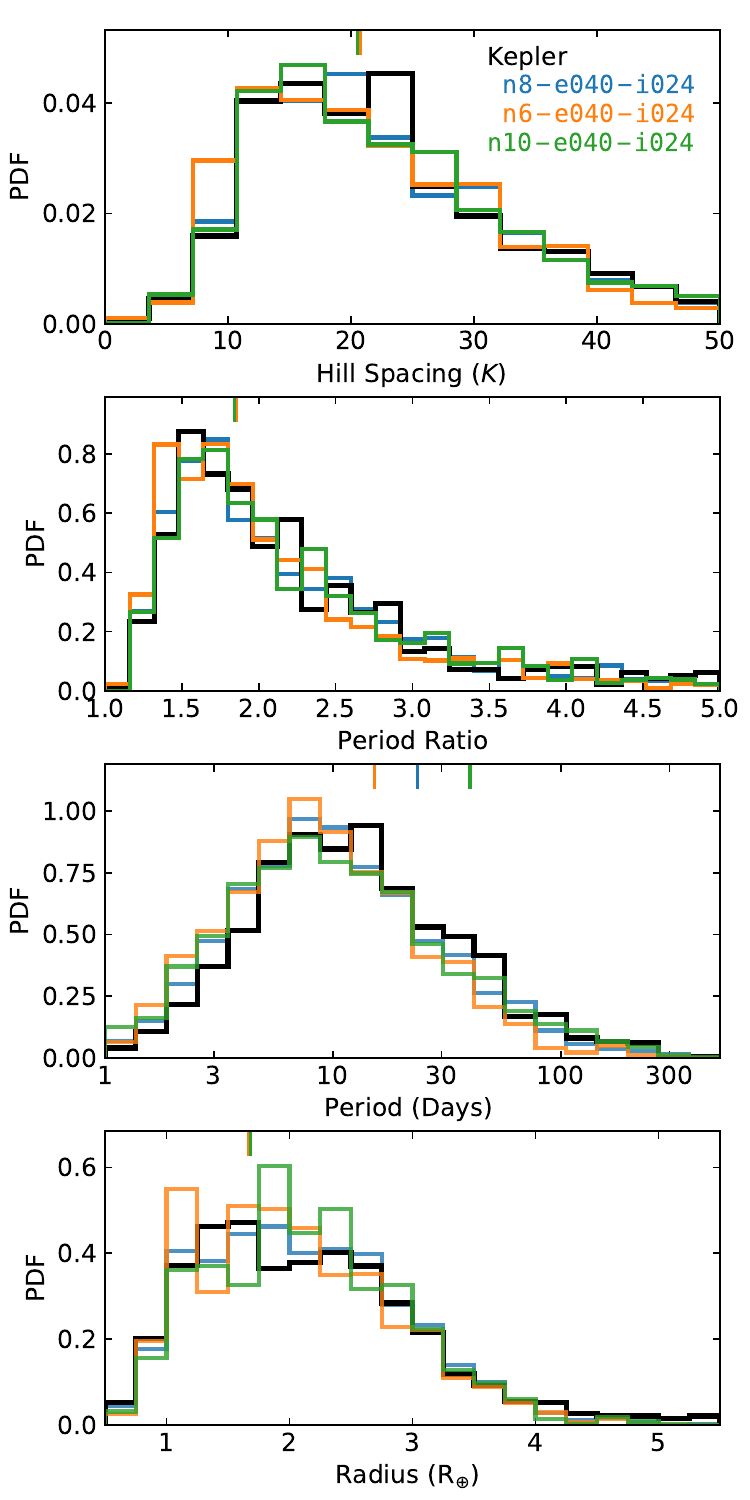}
\caption{Similar to \autoref{fig:dists-eccs} but for ensembles \texttt{n6-e040-i024} (orange), \texttt{n8-e040-i024} (blue) and \texttt{n10-e040-i024} (green).}
\label{fig:dists-nps}
\end{figure}

As expected, systems with higher $\Npi$ become unstable at a shorter $\tau$ and the ensemble reaches the desired dynamical state earlier (\autoref{tab:initial_props}). When the model-detected $K$ distributions agree with the observed, the other properties are also in agreement independent of $\Npi$ (\autoref{fig:dists-nps}). Although the model-detected populations show excellent agreement, the intrinsic $\porb$ distributions are quite different between these ensembles. The lower-$\Npi$ ensembles produce intrinsic systems with lower $\porb$. This is directly related to our setup (\autoref{subsec:setup/orbits}). Systems with lower $\Npi$ are naturally smaller in extent and contain planets restricted to lower $\porb$. In spite of this, the model-detected $\porb$ distributions are consistent, independent of $\Npi$ (\autoref{tab:results_table}). Note that our lowest-$\Npi$ ensemble \texttt{n6-e040-i024} does not quite reach the desired dynamical state even at $t=10^{9}\Pone$. For this ensemble, model-detected distributions for $K$, period ratios, and $\porb$ peak at somewhat lower values when compared to the \kepler\ data (\autoref{tab:results_table}). We believe, had we integrated \texttt{n6-e040-i024} longer, this ensemble also would have converged to similar distributions in these properties.

We also find that the final $\ecc$ and $\imut$ distributions peak at higher values for ensembles with higher $\Npi$, despite having identical initial distributions (\autoref{tab:results_table}), indicating that the orbits in the systems with a higher $\Npi$ get more churned. In our simulations, the intrinsic planet multiplicities can go up to $\Npi$. However, the model-detected populations contain very few ($ < 0.07\%$) systems with $\Np > 6$, irrespective of $\Npi$. Ensembles with higher $\Npi$ have higher fractions of detectable planetary systems with  $\Np \geq 4$  (\autoref{tab:multiplicity}). 

Independent of all the variations in initial properties, none of the ensembles could produce the observed apparent excess of single-transiting planets.

\section{Summary And Conclusions} \label{sec:summary}

In this study, we have investigated the role of dynamical instabilities in the final stages of assembly for planetary systems initially in non-resonant configurations. Starting from a power-law $K$ distribution we simulate our planetary systems with varying initial $\bar{\ecc}$, $\bar{\inc}$, and $\Npi$ until the model-detected systems in each ensemble reach the observed dynamical state, equivalently, $K$ distribution for the \kepler\ multis. Note that this stopping criteria is more stringent and physically motivated compared to the typically used criteria of stopping after a fixed number of orbits when further interactions slow down (\autoref{appendix:stopping}). Also, note that because of our dynamically motivated stopping criteria, the match between the model-detected and observed $K$ distributions should not be considered as validation of models. On the other hand, the fact that dynamical instabilities can produce the same $K$ distribution as observed from a large variety of initial conditions, indicates that the final distributions of properties are rather insensitive to the details of the initial conditions, thus eliminates any need of fine tuning. By design, we do not inject any inter or intra-system correlations at the beginning to find what emerges solely from dynamical encounters. Below we summarize the main findings.
\begin{itemize}
  \item When the dynamical states of our model-detected populations match that of the observed multis, distributions of all other key observable properties agree as well (e.g., \autoref{fig:fidu_dists}). This agreement does not strongly depend on the initial distributions of orbital properties or $\Npi$ (\autoref{fig:dists-eccs}, \ref{fig:dists-incs}, \ref{fig:dists-nps}).
  \item Without any initial input from formation scenarios, host-star dependence, or intra-system correlations, all model ensembles naturally evolve to produce planetary systems very similar in properties to the observed. 
  \item Starting from completely uncorrelated planets, dynamical evolution naturally induces intra-system uniformity in $\Mp$ in the model-detected populations similar to the observed `peas-in-a-pod' trend (\autoref{subsec:res/peas_in_a_pod}, \autoref{fig:m_gini_dist}). The \kepler\ multis seem to favor an initially less correlated population attaining intra-system uniformity through dynamics (\autoref{fig:uniformity_vs_multiplicity}).
  \item The anti-correlation between NAMD and multiplicity manifests as a natural consequence of dynamical instabilities if the initial NAMD is sufficiently low. In this scenario, present-day systems with higher $\Np$ have suffered less dynamical instability in the past.
  \item This NAMD-multiplicity trend, however, disappears in ensembles where the initial NAMD is so high that even the less dynamically evolved high-multiplicity systems retain the initial high NAMD. So if the inferred NAMD-multiplicity anti-correlation in \kepler\ multis is real, pre-instability \kepler\ planets likely had low orbital excitation. On the other hand, the initial NAMD may be constrained from that observed in present-day high-multiplicity systems.
  \item The multiplicity distributions in model-detected populations are similar to those observed for $\Np>2$. However, the observed systems show an excess of singles. This discrepancy cannot be explained simply by varying the initial orbital properties or $\Npi$ within our setup. The observed excess can be explained if $f\approx0.48$ single planets are added to the mix. 
  \item Likely related to the previous point, although, the model-detected $\bar{\ecc}$ for multis is similar to those observed, the model-detected singles have significantly lower $\bar{\ecc}$ compared to the observed singles. This indicates the necessity of an additional population of intrinsically singles, or systems containing higher-mass perturbers which can create dynamically hotter inner planetary systems of sub-Neptunes \citep[e.g.,][]{2017_Hansen,2017_Lai_Pu,2018_Pu_Lai, 2023_Bitsch}.       
\end{itemize}

The success of our model in reproducing several key trends of the observed multis, despite our initial setup was intentionally devoid of any such correlations, indicates that many of the famous observed trends and correlations may have resulted from dynamical processes post-formation and does not point towards specific formation scenarios.

\section*{Acknowledgements}
We thank the anonymous referee for insightful comments and constructive suggestions. TG acknowledges support from TIFR's graduate fellowship. SC acknowledges support from the Department of Atomic Energy, Government of India, under project no.  12-R\&D-TFR-5.02-0200 and RTI 4002. This research has made use of the NASA Exoplanet Archive, which is operated by the California Institute of Technology, under contract with the National Aeronautics and Space Administration under the Exoplanet Exploration Program. All simulations were done on Azure. 

\section*{Data Availability}

The data underlying this article will be shared on reasonable request to the corresponding author.


\bibliographystyle{mnras}
\bibliography{ref}



\appendix

\section{Integration Stopping Time} \label{appendix:stopping}

Here, we compare our choice of simulation-stopping criteria with the common practice in the literature. Traditionally, the ensembles of planetary systems are integrated for $\sim 10^{x}$ $\Pone$, (the adopted x depends on the computational resources, typical values are 7, 8, and 9 depending on the study). Then usually, if the distributions of some critical properties do not change significantly by extending the integration to a few $\times 10^x\Pone$, the system is considered `settled'. Essentially, it means that the occurrence rate of instabilities has decreased sufficiently. This can be problematic because the stability timescale grows exponentially with the Hill spacing \citep[e.g.,][]{1996Chambers, Chatteerjee_2008, 2017Obertas}. So, if a system is integrated till $10^{x}$ $\Pone$, it may acquire properties consistent with a stability timescale of $\sim10^{x}$ $\Pone$. But, the next significant change is expected only at $\sim 10^{x+1}$ $\Pone$. So, checking for changes at a few $\times 10^{x}$ $\Pone$ may not be indicative. It is also not clear when to stop. Since the systems integrated to $10^{x+1}$ $\Pone$ may again become unstable if integrated to say, $10^{x+2}$ $\Pone$, and so on. One natural point to stop would be when only two planets remain. Then we can use a suitable analytic stability criteria \citep[e.g.,][]{GLADMAN1993} to determine the possibility of orbit crossing. Else, a natural stopping point would be the system age. Numerical integration for a large number of systems till their actual ages is computationally impractical. 

\begin{figure}
\includegraphics[width=\columnwidth]{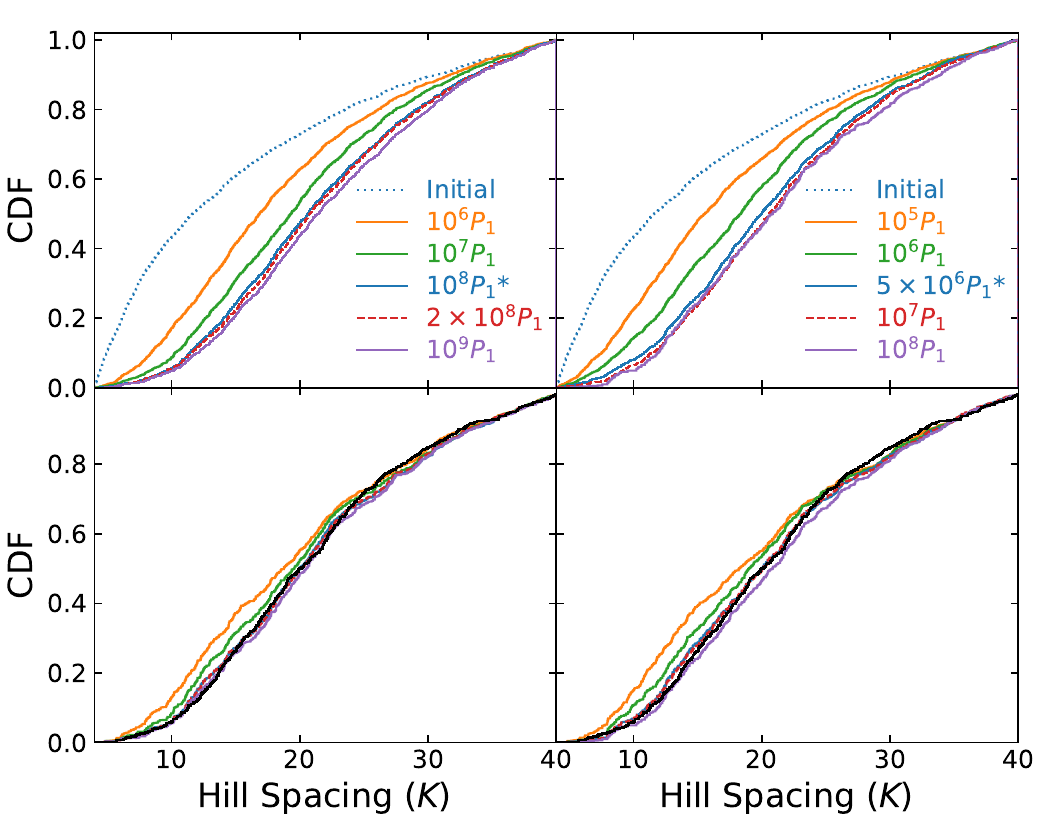}
\caption{CDF for $K$ at different snapshots in time for our fiducial ensemble \texttt{n8-e040-i024} (left) and \texttt{n8-e060-i024}, an ensemble with high initial $\ecc$ (right). Top and bottom panels show the final intrinsic and model-detected distributions. Different colors show the $K$ distributions at different times (see legend). Black denotes \kepler's multis. The snapshot at the stopping time determined by our integration stopping criteria (\autoref{subsec:setup/stopping}) is shown in blue and marked by an asterisk in the legend. Longer simulations show no significant difference.}
\label{fig:K_dist_time_evol}
\end{figure}

Instead, our criteria (as described in \autoref{subsec:setup/stopping}) is more physically motivated. The $K$ distribution in a population essentially indicates how stable or unstable the population is. Hence, bringing all model ensembles to the observed $K$ distribution essentially ensures that the model-detected ensembles are dynamically as active or inactive as the observed. This demanding criteria already fulfills the traditional settling-down criteria. For example, our fiducial ensemble (\texttt{n8-e040-i024}) satisfies our stopping criteria earliest at $\sim10^{8}\Pone$. We find no significant changes in the $K$ distribution after further integration (\autoref{fig:K_dist_time_evol}, left panels) to $2\times10^8\Pone$ or even $10^9\Pone$. In fact, changes in the ensemble properties slow down significantly after $\sim10^7\Pone$. On the other hand, different ensembles may reach this dynamical state at different multiples of $\Pone$ based on their initial conditions (\autoref{tab:initial_props}), since the initial conditions dictate the timescale for the onset of instabilities. For example, the ensemble (\texttt{n8-e060-i024}) satisfies our stopping criteria at $5\times10^{6}\Pone$, and we find no significant changes in the Hill spacing distribution, even if we continue the integration for more than an order of magnitude longer (\autoref{fig:K_dist_time_evol}, right panels). In contrast, the ensembles with lower initial orbital excitation require longer simulations (up to $\sim 10^{9}\Pone$ performed in our models) to reach the same dynamical state (\autoref{tab:initial_props}). Thus, using a fixed multiple of $\Pone$ for stopping time would not necessarily mean that the dynamical states of the ensembles, in the end, are similar. Thus, when ensembles with varied initial orbital properties are considered, we argue that all ensembles should be brought to a compatible dynamical state consistent with the observed before comparison. 

\section{Intra-System Uniformity}
\label{appendix:l23}
While preparing our manuscript, we came to know about the results of L23. In this fantastic work, L23 independently reach the same conclusion regarding the intra-system uniformity induced by dynamical instabilities despite a very different initial setup compared to ours. This strengthens the result that dynamical instabilities in multiplanet systems with little intra-system uniformity, independent of the initial setup, can make systems more uniform over time. Here we compare our results with an additional simulated ensemble using the exact initial setup of L23. 

The initial setup of L23 is significantly different from ours. Moreover, unlike ours, the scope of their study is limited to investigating the origins of intra-system uniformity. Hence, we limit our comparisons only to this aspect. For completeness, L23 consider initial planetary systems with $10$ planets orbiting a sun-like star. They assume a flat initial period ratio distribution within $[1.10, 1.50)$, and draw $\Mp$ from a Gaussian distribution centered around $\Mp/M_{\earth}=3$. They use the $\Mp$-$\Rp$ relationship given in \citet{Wolfgang_2016}. We replicate these initial conditions in this additional ensemble. 

L23 find initial (final intrinsic) $\mathcal{D}=0.59$ ($0.48$) after $10^{9} \Pone$. In our simulated ensemble using their initial setup, we also find a similar result, initial (final intrinsic) $\mathcal{D}=0.60$ ($0.48$). L23 have not considered the effects of various detection and sensitivity biases in their analysis. Our results suggest that this can make a difference (\autoref{fig:m_gini_dist}, \autoref{subsec:res/peas_in_a_pod}). When we apply our planet detection algorithm (\autoref{subsec:setup/detections}) to create a \kepler-detectable population from this ensemble, we find a model-detected $\mathcal{D}=0.35$, significantly lower than observed (\autoref{subsec:res/peas_in_a_pod}, \citet{Goldberg_2022}, \citet{lammers2023intrasystem}).

Interestingly, L23 report intra-system uniformity in spacing measured in period ratios. They find that period ratios within a system with $\Np\geq3$ in their intrinsic population are more correlated (correlation coefficient $\mathcal{C}=0.39$) than those overall. We find a similar result, $\mathcal{C}=0.48$, in our intrinsic final ensemble using their initial setup. However, this enhanced intra-system correlation in spacing disappears when transit and detection biases are taken into account. Our model-detected population using their initial setup shows $\mathcal{C}=0.03$. Interestingly, if we restrict our fiducial ensemble to $4<K<20$, roughly equivalent to the range of period ratios considered in L23, we also find significant intra-system uniformity in spacing in the final intrinsic population, $\mathcal{C}=0.21$. However, this uniformity disappears when transit and detection biases are taken into account, $\mathcal{C}=0.03$.

\bsp    
\label{lastpage}
\end{document}